\documentclass[twocolumn,10pt]{IEEEtran}
\usepackage{algorithm,algorithmic,amsbsy,amsmath,amssymb,epsfig,bbm,mathrsfs,fancyhdr,cases,comment,color}

\newcommand{\nn}{\nonumber}
\newcommand{\beq}{\begin{equation}}
\newcommand{\eeq}{\end{equation}}
\newcommand{\beqn}{\begin{eqnarray}}
\newcommand{\eeqn}{\end{eqnarray}}
\DeclareMathOperator{\ex}{\mathbb{E}}

\begin{document}

\title{Distributed and Centralized Hybrid CSMA/CA-TDMA Schemes for Single-Hop Wireless Networks}

\author{Bharat Shrestha, Ekram Hossain, and Kae Won Choi
\thanks{B. Shrestha and E. Hossain are with the Department of Electrical and Computer Engineering, University of Manitoba, Canada. K. W. Choi is with Seoul National University of Science and Technology, Korea. This work was supported by the Natural Sciences and Engineering research Council of Canada (NSERC) Discovery Grant 249500-2009. 
}



\vspace{-5mm}
}

\maketitle

\begin{abstract}
The  strength of carrier-sense multiple access with collision avoidance (CSMA/CA) can be combined with that of time-division multiple access (TDMA) to enhance the channel access performance in wireless networks such as the  IEEE 802.15.4-based wireless personal area networks (WPANs).
In particular, the performance of legacy CSMA/CA-based medium access control (MAC) scheme in congested networks can be enhanced through a hybrid CSMA/CA-TDMA scheme while preserving the scalability property. In this paper, we present distributed and centralized channel access models which follow the transmission strategies based on Markov decision process (MDP) to access both contention period and contention-free period in an intelligent way. The models consider the buffer status as an indication of congestion provided that the offered traffic does not exceed the channel capacity. We extend the models to consider the hidden node collision problem encountered due to the signal attenuation caused by channel fading. The simulation results show that the MDP-based distributed channel access scheme outperforms the legacy slotted CSMA/CA scheme. This scheme also works efficiently in a network consisting of heterogeneous nodes. The centralized model outperforms the distributed model but requires the global information of the network.


\end{abstract}

\section{Introduction}
Hybrid carrier-sense multiple access with collision avoidance (CSMA/CA) and time-division multiple access (TDMA) protocols such as the IEEE 802.15.4 standard-based medium access control (MAC) protocol~\cite{stand} are useful in realizing low-power and low-rate wireless networks. TDMA is a collision-free channel access mechanism whereas CSMA/CA is a contention-based MAC protocol. TDMA is desirable to reduce collisions and to conserve power for channel access. However, CSMA/CA could be used by the wireless nodes to send the channel access request in a hybrid CSMA/CA-TDMA-based wireless network. In such a network operating in the beacon-enabled mode (e.g.,  IEEE 802.15.4 network), wireless nodes synchronize their superframes with the coordinator by the help of a beacon frame.

The CSMA/CA operation requires a node to perform carrier sensing to make sure that the channel is free for transmission. The node competes with other nodes during contention access period (CAP) to get access to the channel and transmit packets to the coordinator using the CSMA/CA mechanism. On the other hand, a node can transmit packets in a collision-free manner using TDMA slots during contention-free period (CFP) without using any carrier-sensing mechanism. Whenever a node requires a certain guaranteed bandwidth for transmission, the node sends a reservation request for TDMA slot by using CSMA/CA during CAP. Upon receiving the request, the coordinator first checks the availability of the TDMA slots and it informs the node of the allocation of the TDMA slot. When a TDMA slot is allocated, the node can turn off its receiver circuitry during CAP and go to low power mode to save its limited battery power. Transmission using TDMA slot also reduces congestion during CAP. Although reservation-based TDMA  provides collision-free transmission, a node has to transmit the  reservation request successfully during CAP. Some disadvantages of using only TDMA slot-based transmissions are: i) due to the fixed frame length, the packet transmission delay increases with increasing frame length (i.e., beacon interval), ii) the channel is under-utilized when traffic demand is low, and iii) when traffic demand is high, there is only fixed amount of allocated bandwidth (or limited number of TDMA slots).

Transmissions using CSMA/CA during CAP can avoid some of the above mentioned problems of the TDMA slot-based transmissions; however, only with CAP, packet transmission requirements (e.g., throughput, energy efficiency) may not be satisfied, especially when the network is congested.
For transmissions during CAP, since the nodes compete with each other to get access to the channel, the network gets congested  as the network size grows. Congestion drives the network into saturation worsening the performance in terms of latency and energy-consumption. In such a scenario, the nodes may have to take multiple backoffs before attempting their transmissions. Congestion may occur during CAP  even when the total packet arrival rate into the network does not exceed the flow capacity of the contention period. Hidden node collision, which  is a common problem in CSMA/CA-based wireless networks, also affects packet transmissions during CAP. An increased number of collisions results in an increase in the number of retransmissions and hence leads to reduced packet service rate. In a similar manner, signal attenuation due to channel fading as well as interference may lead to increased number of retransmissions in the network.  The channel access scheme in the network should therefore be able to adapt to the network dynamics and perform efficiently in congestion scenarios. In particular, dynamic switching between the transmission modes using CAP and CFP would be desirable to achieve a superior channel access performance~\cite{hybrid}.

In this work, we model and analyze distributed and centralized channel access schemes that use both contention and contention-free accesses to cope with the above mentioned problems. For both of these schemes, to determine the strategy for data transmissions during a superframe, we formulate Markov Decision Process (MDP)~\cite{puterman} models to decide whether to {\em transmit using contention period}, or {\em transmit using contention free period}, or {\em both}, or {\em not to transmit at all}. This work provides a method of changing the legacy CSMA/CA scheme to a hybrid CSMA/CA-TDMA scheme and improving the channel access performance of the nodes while preserving the scalability property of CSMA/CA. The novelty of the proposed channel access schemes is that they incorporate the notion of optimality in channel access considering the properties of both CSMA/CA and TDMA. \textcolor{red}{The main objective of this paper is to improve the performance of hybrid CSMA/CA-TDMA channel access schemes in terms of energy efficiency and throughput. Note that the performance gain is achieved at the cost of added computational complexity in the system.}
The main contributions of this work can be summarized as follows:

\begin{itemize}

\item For low-power CSMA/CA-TDMA hybrid MAC protocol, we develop an MDP-based Distributed Channel Access (MDCA) scheme,   which considers both the throughput and the energy consumption of the wireless nodes.   In this scheme, a node is unaware of the traffic loads of the other nodes in the network and the coordinator does not require any information from the nodes. This scheme provides an improved TDMA slot utilization over the scheme proposed in~\cite{bharatglb}.

\item We develop an MDP-based Centralized Channel Access (MCCA) scheme, which improves the energy consumption rate compared to the existing hybrid CSMA/CA-TDMA schemes. However, it requires the traffic information of all the nodes available at the central controller and more computational efforts. 

\item We extend the models to consider the effect of channel fading.

\item We provide a comprehensive performance evaluation of the proposed channel access schemes and a comparison with the traditional CSMA/CA-based channel access schemes as well as two other hybrid CSMA/CA-TDMA schemes in the literature.

\end{itemize}

In Section II, we describe the system model and assumptions and also introduce the proposed MDP-based MAC schemes. In Section III, we formulate the MDP problem for the distributed channel access scheme. In Section IV, we present the MDP-based centralized channel access scheme. We analyze the effect of hidden node collision in Section V. In Section VI, we present the performance evaluation results for the proposed channel access schemes. We discuss the related work in Section VII. Section VIII draws the conclusion. Table~\ref{symbolsused} lists the major notations used in this paper.

\begin{table*}[t]
\caption{List of notations}
\centering
\begin{tabular}{|l|l|l|l|}
\hline \hline
Notation & Meaning & Notation & Meaning\\
\hline \hline
$N$ & Number of nodes & $\alpha_{n|N}$ & Probability of channel being idle during first carrier   \\
                          & &  & sensing for node $n$ given $N$ competing nodes during CAP  \\ \hline
 $A_{t,n}$  & Action of node $n$ at superframe $t$  & $\beta_{n|N}$ & Probability of channel being idle during second carrier    \\
                       & &   & sensing for node $n$ given $N$ competing nodes during CAP  \\
                       & &   & and the channel was idle during  first carrier sensing   \\ \hline
 $\Lambda$ & Set of all possible actions  & $\Phi_{n|N}$ & Throughput of a node $n$ given $N$ competing nodes \\
                       & &  &  during CAP  \\ \hline
 $S_{t,n}$ & State of node $n$ at superframe $t$  & $\Phi_{cap}$ & Total number of packets retrieved out of MAC buffer \\
                       & &   & during CAP  \\ \hline
 $B_{t,n}$ & Buffer state of node $n$  & $ \kappa$ & Total number of packets successfully transmitted to \\
                  & at superframe $t$  &   & the coordinator during CAP  \\ \hline
 $\mathbf{B}_t$ & Joint buffer state of $N$ nodes  & $P_c$ & Probability of collision  \\
  &  $(B_{t,1}, B_{t,2}, \cdots, B_{t,N})$   & &  \\
\hline
 $B_{max,n}$ & Maximum  buffer size  of node $n$ & $\eta$ & Number of packets that can be transmitted  \\
           &   &   & during a slot duration \\ \hline
 $R_{s,a}$ & Reward when action $a$ is taken & $\Theta$ & Probability of outage \\
     & at state $s$  & & \\ \hline
 $\gamma$ & Discount factor  & $H$ & Hidden node collision probability \\ \hline
 $\lambda$ & Average packet arrival rate  & $\Psi_n$ & Set of nodes which are hidden to node $n$ \\ \hline
 $T_{sf}$ & Length of a superframe   & $T_{cap}$ & Length of CAP   \\ \hline
 $T_{cfp}$ & Length of CFP    & $T_{slot}$ & Length of a slot  \\ \hline

\hline
\end{tabular}
\label{symbolsused}
\end{table*}

\section{System Model, Assumptions, and the Hybrid CSMA/CA-TDMA Schemes}

\subsection{Network Model}

\begin{figure}[t]
\begin{center}
\includegraphics[width = 3.5in]{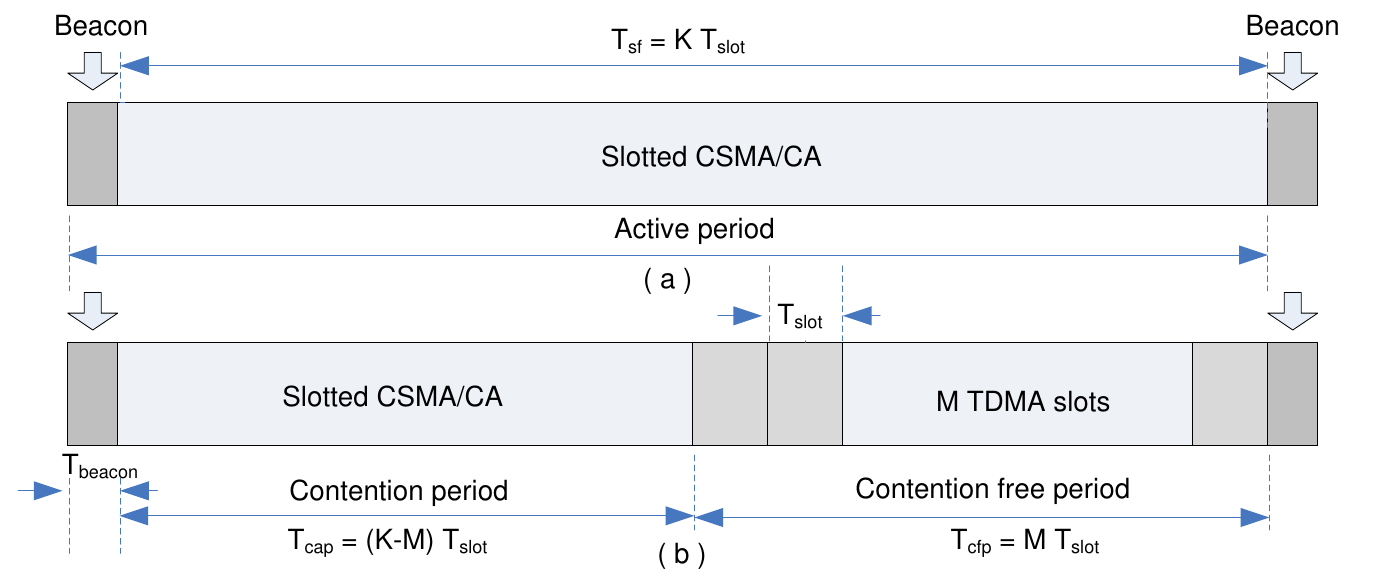}
\caption{Superframe structure: (a) with CFP length = 0, (b) with CFP length = $M$ slots.}
\label{structure}
\end{center}
\end{figure}

We consider a star network topology with $N$ nodes and a network coordinator. A node is indexed by $n$ $(= 1,\cdots,N)$. Each node is within the carrier-sensing range of the other nodes when statistical variation in the channel propagation condition is not considered. Time is divided into superframes each of which has a contention period of $T_{sf}$ unit backoff period (UBP) plus a beacon frame of length $T_{beacon}$ UBP. The contention period (i.e., the superframe duration excluding beacon frame) is divided into $K$ slots and the length of each slot is $T_{slot}$ UBP (i.e., $T_{sf} = K T_{slot}$) as shown in~Fig.~\ref{structure}. A node can transmit $\eta$ packets during a TDMA slot.
The superframe structure, which is similar to the standard IEEE 802.15.4 superframe structure, is shown in Fig.~\ref{structure}. The coordinator uses $T_{beacon}$ UBP of the superframe to broadcast the beacon frame. Let $M$ be the number of slots during CFP. The length of the CAP available for the nodes is $T_{cap} = T_{sf}-MT_{slot}$ UBP and the length of CFP is $T_{cfp} = M T_{slot}$ UBP.
A node uses the slotted CSMA/CA protocol to access the medium during the CAP, while a node can only transmit on an assigned TDMA slot during the CFP. We refer to~\cite{stand} for the detailed operation of the slotted CSMA/CA.
Note that when $M = 0$, nodes use only CSMA/CA for data transmissions. When $M = K$, the length of CAP is zero and the nodes transmit only in the assigned TDMA slots. In addition to CAP and CFP, an inactivity period can also be introduced in the superframe as in the IEEE 802.15.4 MAC. \textcolor{red}{Note that with introduction of longer inactivity periods and multiple coordinators, this model can be extended to multihop networks. For this, multiple coordinators can be synchronized such  that the beacon frames and active periods of the different coordinators are non-overlapping.} However, in this paper, we focus only on single-hop networks and consider only the active period in a superframe $t$ which starts at $(T_{sf}+T_{beacon})t$.

\subsection{Traffic Model}
In this section, we explain the traffic model for a node. Each node has a packet buffer. The maximum number of packets that are stored in the buffer of node $n$ for transmission is denoted by $B_{max,n}$. As described in~\cite{bharatlet}, we consider the batch Poisson process as the packet arrival model for a node. \textcolor{red}{We assume that packets are queued in MAC buffer and are not transmitted until the start of the next superframe. While queuing during a superframe, the number of packets in the buffer might be greater than $B_{max}$. In this case, all the packets which arrived later than first $B_{max}$ packets in the buffer are discarded at the end of the superframe.} 
The arrival time and the number of packets of the $j^{th}$ batch at node $n$ are denoted by $Y_{n,j}$ and $Z_{n,j}$, respectively.
For node $n$, the inter-arrival time between batches, $Y_{n,j+1}-Y_{n,j}$, is exponentially distributed with mean $1/\lambda_n$.
The number of packets in each batch, $Z_{n,j}$, is identically and independently distributed.
The probability mass function (pmf) of $Z_{n,j}$ is denoted by $f^Z_n$.

We denote by $X_{n,\Gamma}$ the number of arrived packets at node $n$ during a time interval the length of which is $\Gamma$.
Since the packet arrivals follow a batch Poisson process, $X_{n,\Gamma}$ follows the compound Poisson distribution \cite{Adelson}.
Therefore, the characteristic function of $X_{n,\Gamma}$ for $\tau \in \mathbb{R}$ is given as
\begin{align}
\varphi^X_{n,\Gamma}(\tau) = \ex[\exp(i \tau X_{n,\Gamma})] =\exp\{\lambda_n\Gamma(\varphi^Z_n(\tau)-1)\}
\end{align}
where $\varphi^Z_n(\tau)$ is the characteristic function of $Z_{n,j}$ such that $\varphi^Z_n(\tau)=\ex[\exp(i \tau Z_{n,j})]=\sum_z \exp(i\tau z) f^Z_n(z)$.
The pmf of $X_{n,\Gamma}$, denoted by $f^X_{n,\Gamma}$, can be derived from $\varphi^X_{n,\Gamma}$ by using the inverse formula for the characteristic function.

\subsection{Node Actions for Channel Access}
We define $\pi^*_{N,K,M}$ as the transmission policy for each node when the network size is $N$, the superframe length is $K$ slots, and the length of CFP is $M$ slots ($T_{cap} = MT_{slot}$UBP). Then length of CAP is $K-M$ slots ($T_{cap} = (K-M)T_{slot}$ UBP).\footnote{\textcolor{red}{Depending on the value of $N$, the coordinator can adjust the values of $K$ and $M$ periodically with a view to improving the system performance. Optimization of the values of $K$ and $M$ is however out of the scope of the paper.}}
The policy can be determined by solving the MDP problem to be described later in this paper.
The policy $\pi^*_{N,K,M}$ maps the current state (i.e., the current buffer level $B$) to an action $A$ (i.e. $B\rightarrow A$).
According to the policy, in each superframe, based on its current packet buffer level, a node selects an action out of the following four actions: {\em defer transmission ($a_1$}), {\em transmit packet during CAP ($a_2$}), {\em transmit packet during CFP ($a_3$)}, and {\em transmit packet during both CAP and CFP ($a_4$)}.

\subsubsection{MDCA scheme}
Table~\ref{MDCA_scheme} shows the operation of nodes in the MDCA scheme. In this scheme, the coordinator divides the superframe into a fixed-size CFP ($M$ slots) and a fixed-size CAP ($K-M$ slots). Each node receives a beacon at the beginning of the superframe $t$ and obtains information such as the network size $N$, the length of CAP ($K-M$ slots), and the length of CFP ($M$ slots). Note that some or all of the $M$ slots in CFP might be occupied or empty.
From this information, each node distributedly determines the policy $\pi^*_{N,K,M}$.
Let $G_{t,n}$ denote a TDMA slot indicator for node $n$ in superframe $t$.
If node $n$ is allocated a slot in superframe $t$, we have $G_{t,n} = 1$; otherwise $G_{t,n} = 0$.
According to the policy $\pi^*_{N,K,M}$ and the TDMA slot indicator $G_{t,n}$, node $n$ performs the  operations (i.e., $(B,G)\rightarrow A$) as described below.

If $A_{t,n} = a_1$ (defer transmission) and $G_{t,n} = 0$, node $n$ does nothing but waits for the next beacon frame.
If $A_{t,n} = a_2$ (transmit packet during CAP) and $G_{t,n} = 0$, node $n$ tries to transmit packets by using slotted CSMA/CA during the CAP in superframe $t$.
If there is not enough time to transmit a packet during the current CAP, node $n$ waits until the next beacon frame. 
In the case that $A_{t,n} = a_1$ (defer transmission) or $A_{t,n} = a_2$ (transmit packet during CAP) when $G_{t,n} = 1$, node $n$ has to empty the slot by sending a packet with the TDMA slot de-allocation request bit set during allocated time slot in CFP.

If $A_{t,n} = a_3$ (transmit packet during CFP) and $G_{t,n} = 1$, it transmits only in the assigned TDMA slot.
If no TDMA slot has been assigned to node $n$ ($G_{t,n} = 0$), in the case that $A_{t,n}=a_3$, node $n$ sets the TDMA slot request bit in the data packet and transmits the packet by using the slotted CSMA/CA in the CAP.
If at least one TDMA slot among $M$ slots is available, the coordinator assigns a TDMA slot to node $n$ and notifies node $n$ of the assigned slot number in the acknowledgment packet.
If node $n$ is notified of the assigned TDMA slot in the acknowledgment packet, the node halts transmission during the CAP and resumes  transmission in the assigned slot during CFP in the same superframe.
Otherwise, the node continues to transmit using the slotted CSMA/CA scheme as long as there is enough time left in the CAP.

If $A_{t,n} = a_4$ (transmit packet during both CAP and CFP), $G_{t,n} = 1$ and $B_{t,n} = b$, node $n$ attempts to transmit $\max(b-\eta,0)$ packets using CSMA/CA during CAP and transmits $\min(\eta,b)$ packets during the assigned TDMA slot in the CFP. If a slot has not been assigned ($G_{t,n} = 0$), node $n$ follows a procedure similar to that for action $A_{t,n} = a_3$ to send the TDMA slot request.

Note that the MDCA scheme requires a contention period which is  long enough to send the request successfully.
If $T_{tx}$ denotes the packet transmission time including acknowledgment, inter-frame space, and propagation time, then for the MDCA scheme, the length of the contention period should be at least $NT_{tx}$. To prevent the starvation of other nodes in accessing the TDMA slots, a node leaves the assigned TDMA slot after using it for a predefined number ($\varrho$) of consecutive superframes. Since the policy is developed offline, complexity is not a big issue for the nodes.
\begin{table}[t]
\scriptsize
\caption{Node Actions in the MDCA scheme}
\centering
\begin{tabular}{|l|l|l|}
\hline \hline
 & $G_{t,n} = 0$ & $G_{t,n} = 1$ \\
\hline \hline
$A_{t,n} = a_1$ & Do nothing & Set slot de-allocation request bit \\
  &  & in the transmitted packet  \\ \hline
$A_{t,n} = a_2$ & Transmit during CAP & Set slot de-allocation request bit \\
  &  & in the transmitted  packet  \\ \hline
$A_{t,n} = a_3$ & Set slot request bit  & Transmit during CFP  \\
                 &     in the transmitted packet   & \\ \hline
$A_{t,n} = a_4$ & Set slot request bit  & Transmit during CAP and CFP  \\
               &  in the transmitted packet  & \\ \hline
\hline
\end{tabular}
\label{MDCA_scheme}
\end{table}

\subsubsection{MCCA scheme}
\begin{figure}[t]
\begin{center}
\includegraphics[width = 3.5in]{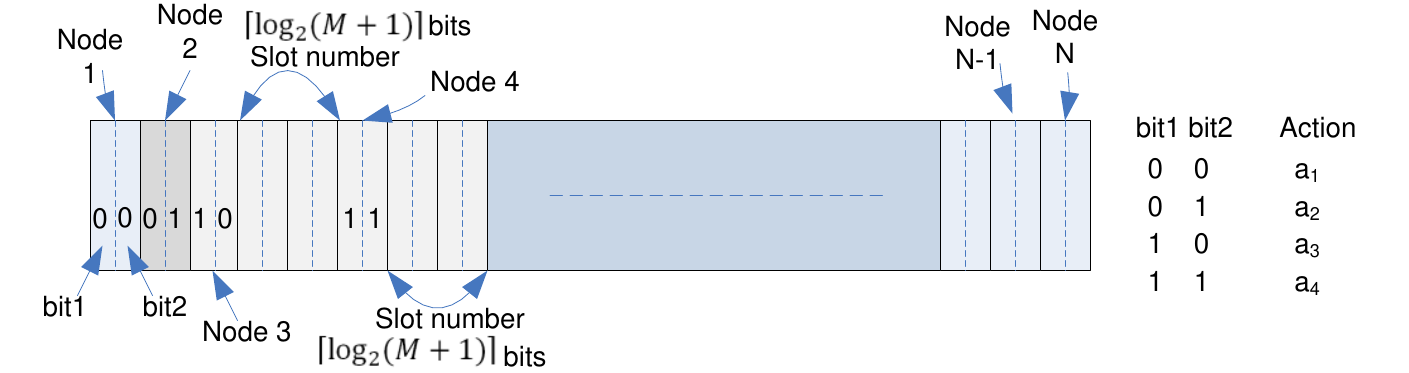}
\caption{Format of actions and TDMA slot numbers of $N$ nodes in the MCCA scheme.}
\label{s_list}
\end{center}
\end{figure}

In this scheme, the coordinator divides the superframe into a CFP the length of which is $M (0\leq M \leq M_{max})$ slots, and a CAP the length of which is $K-M$ slots. Note that $M$ slots are allocated to the needy nodes according to a policy and $M_{max}$ is the maximum number of slots available for CFP in this case. With the MCCA scheme, it is assumed that the coordinator has the information of packet arrival rates and buffer levels of all the nodes associated with it. The information of the packet arrival rate can be sent to the coordinator during the node association phase. The coordinator receives the value of the buffer level each time a data packet is received from the node because the information of buffer level is piggybacked by the data packet. For given $K$ and $M$, the coordinator determines the transmission policy $\pi^*_{N,K,M}$ for each node to reduce the overall energy consumption. For observed buffer level $\textbf{B}_t$ of $N$ nodes, the coordinator then broadcasts the policy (i.e., action to be taken by each node ($\textbf{B} \rightarrow \textbf{A}$)) through the beacon frame. The beacon frame includes the list of actions for $N$ nodes in a format shown in Fig.~\ref{s_list}. The actions $a_3$ and $a_4$ are followed by the TDMA slot numbers. Although this scheme can provide a better performance than the MDCA scheme, the complexity grows exponentially with the network size. Therefore, for this approach, we propose an approximate solution to find the transmission policies.

\subsection{Beacon Loss and Change in Network Size}

When a node misses the beacon frame in a superframe $t$, in the case of MDCA scheme, the node calculates the lengths of the CAP and CFP from the last received beacon frame and uses the CSMA/CA scheme to transmit packets during the CAP. If collision occurs more than once, the node waits for the next beacon frame. \textcolor{red}{The reason for this is that the collision might have occurred during a time slot allocated to other nodes.} The node also uses the slotted CSMA/CA to transmit packets during the assigned TDMA slot in the CFP if the node has not sent any de-allocation request in the last superframe. In case of a collision, the node waits for the next beacon frame. However, with the MCCA scheme, since the node will miss the transmission policy broadcast from the coordinator, it will attempt to access the channel during the CAP. This might cause increased congestion during the CAP and/or wastage of the TDMA slot in superframe $t$ in case the policy has been changed. Throughout this paper we assume that there is no beacon loss in the network.

When a node joins or leaves the network (e.g.,  network consisting of energy harvesting sensor nodes or mobile nodes), the network coordinator updates the size of network $N$. For example, a node can be considered dead if the coordinator does not receive any packets from the node for a predefined number of consecutive superframes. A new node sends the association request to the coordinator using slotted CSMA/CA during CAP. In the MDCA scheme, a node determines the policy $\pi^*_{N,K,M}$ based on $N$. Note that $N$ is obtained through the beacon frame. In the case of the MCCA scheme, the coordinator takes into account the current network size $N$ to determine the transmission policy.

\subsection{An Analytical Model for Slotted CSMA/CA}

In the design of MDCA and MCCA schemes, the throughput in saturation mode is taken into account because each node assumes that the other $N-1$ nodes have packets to transmit during the superframe period.  Therefore, in this section, we  calculate the throughput ($\Phi_{cap}$) of the nodes during CAP by including the probability of channel outage ($\Theta$) which induces congestion in the network~\cite{bharattitb}. During the CAP, each node in the network uses slotted CSMA/CA as defined in the IEEE 802.15.4 standard-based MAC protocol~\cite{stand}. The parameters, namely,  $\alpha$ (i.e., the probability of channel being idle during first carrier sensing), $\beta$ (i.e., the probability of the channel being idle during second carrier sensing given that the channel was idle during first carrier sensing), and $\Phi_{cap}$ (i.e., MAC throughput) depend on the congestion in the network (e.g., the number of nodes $N$ in the network and the CAP length which is $T_{cap}$ UBP). We refer to~\cite{patro},~\cite{park1} for the details of solving a discrete-time Markov chain model and finding the parameters in the saturation mode (i.e., when all the nodes have packets to transmit). We treat retransmission due to collision same as retransmission due to outage. Taking the effects of channel outage into account, the probability of collision $(\widetilde{P}_c)$ is updated as follows:
\beqn
 \widetilde{P}_c = P_c(1-\Theta) + \Theta. \label{pctilda}
\eeqn
We solve the discrete-time Markov chain model using the  probability of collision in (\ref{pctilda}). We define $\hat{P}_{cs}$ as the virtual probability of carrier-sensing due to outage probability as follows:
\beqn
\hat{P}_{cs} = 1 - (1 - \widetilde{P}_c)^{\frac{1}{N-1}}.
\eeqn
\textcolor{red}{Note that the derivation is based on the expression for probability of collision given by: $P_c = 1 - (1-P_{cs})^{N-1}$.} Then, the MAC goodput ($\kappa$) is expressed as the probability that no other nodes start carrier sensing as follows:
\beq
 \kappa = \alpha \beta P_{cs} (1 - \hat{P}_{cs})^{N-1}
\eeq
where the probability of carrier-sensing ($P_{cs}$) is determined by solving the discrete-time Markov chain model. As defined in~\cite{park1}, the probability of packets being discarded due to the limit on the maximum number of backoff  ($P_{discard}$) is given as
\beqn
  P_{discard} = \phi^{m+1} \frac{1 - (\widetilde{P}_c (1 - \phi^{m+1}))^{W+1}}{1 - \widetilde{P}_c (1 - \phi^{m+1})}
\eeqn
in which $m$ is the maximum number of backoffs allowed for a transmission, $W$ is the maximum number of retransmissions allowed before a packet is dropped, and $\phi = (1 - \alpha \beta)(1 - P_d)$ is the probability of going to another backoff stage due to channel being busy given that the packet is not deferred. A packet is deferred when there is not enough time left in the current superframe to transmit a packet. The probability that transmission of a packet is deferred is $P_d = \frac{T_{tx}}{T_{cap}}$, where $T_{tx}$ is the packet length (in time) including time to receive acknowledgment packet and propagation time. The probability of packet dropping due to maximum number of retransmission  ($P_{drop}$) is simply $P_{drop} = \widetilde{P}_c^{W+1}$.
Then the MAC throughput $\Phi_{cap}$, \textcolor{red}{taking into account the discarded packets and the dropped packets}, is estimated as
\beq \label{threshx}
 \Phi_{cap} = \frac{\kappa}{(1-P_{discard})(1-P_{drop})} T_{cap}
\eeq
where $T_{cap}$ is the contention access period in terms of number of backoff units.

\begin{figure}[t]
\begin{center}
\includegraphics[width = 3.5in]{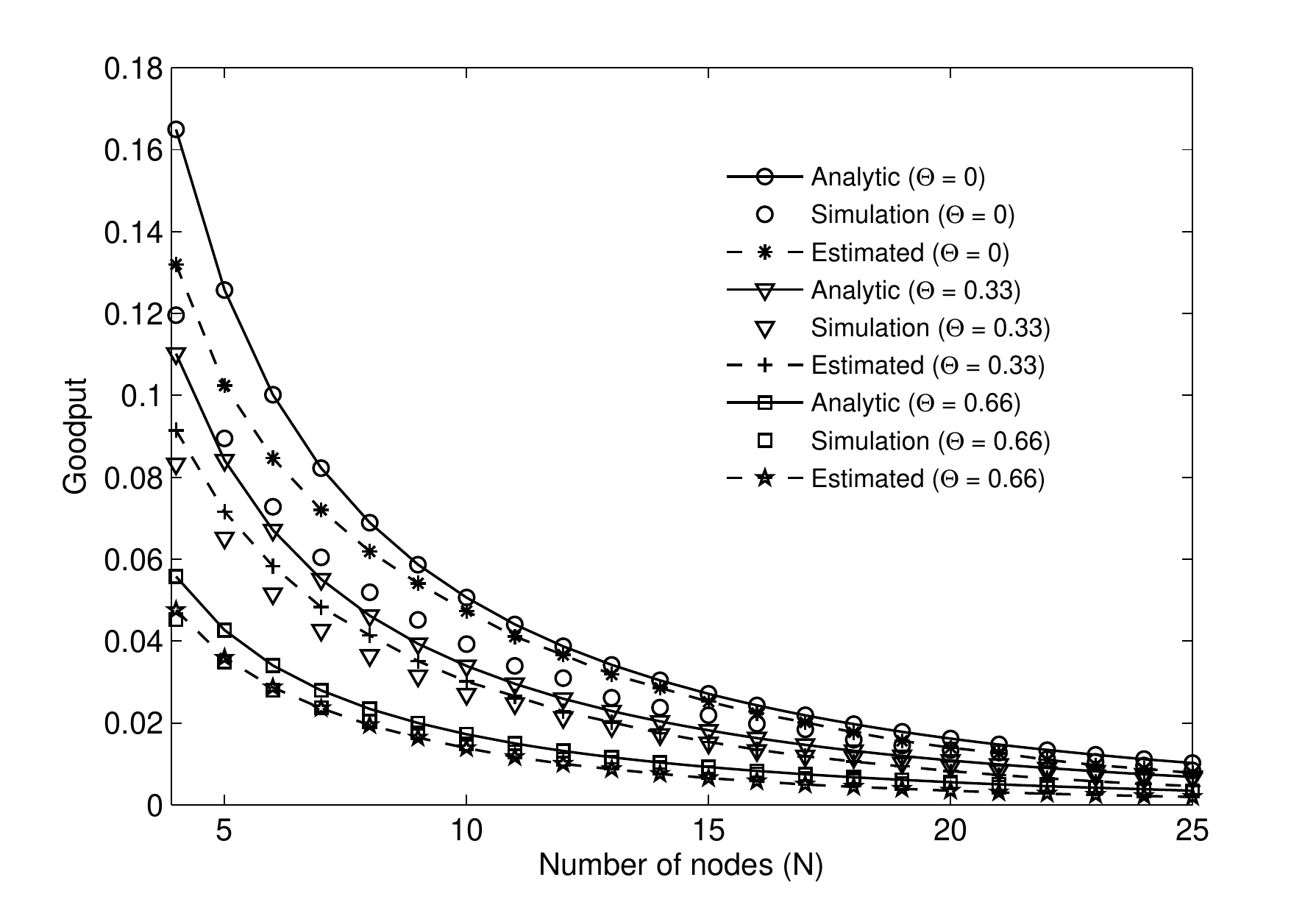}
\caption{Saturation throughput for different values of channel outage probabilities.}
\label{thro_outage}
\end{center}
\end{figure}

By  MATLAB simulations, we observe the variation in throughput of the hybrid MAC in the beacon-enabled mode with respect to the probability of channel outage. In these simulations, we assume superframe length of $T_{sf} = 384$ unit backoff period (UBP) and zero inactive and contention-free periods. We consider packet length $T_{tx} = 10$ UBP including time for acknowledgment and propagation time. We assume that the nodes start random backoff before starting carrier sensing. In the simulation, to determine $\Phi_{cap}$, we count the average number of packets per superframe that the nodes accept at the MAC layer. Then we calculate goodput $\kappa = \Phi_{cap}(1-P_{discard})(1-P_{drop})/T_{sf}$, which is marked as `Estimated' in Fig.~\ref{thro_outage}. To estimate goodput $\kappa$ directly, we also count the average number of packets transmitted successfully, which is marked as `Simulation' in the figure. Fig.~\ref{thro_outage} shows that the results on estimated and the analytical throughput during CAP follow the simulation results. The small gap in the curves is due to the deferred transmissions. Note that the lower the number of nodes, the higher is the packet arrival rate in the saturation region and higher is the probability of deferred transmission of each node.

In the case of heterogeneous nodes (i.e., when the traffic and the MAC parameters are non-identical for the different nodes), we calculate the values of $P_{cs,n}$, $\alpha_n$, and $\beta_n, \  \forall n \in \{1, 2, \cdots, N\}$ for the IEEE 802.15.4 MAC using a discrete-time Markov chain model. For the details of the model and the derivations,  we refer to~\cite{patro},~\cite{bharattitb}, and~\cite{park}. The probability of collision $P_{c,n}= 1-\prod_{\substack{j=1 \\ j \neq n}}^N(1-P_{cs,j})$ is the probability that at least one among $N-1$ nodes starts carrier sensing in the CAP. The throughput, when $N$ nodes are active, is given as $\Phi_{n|N} = \alpha_n \beta_n P_{cs,n} \prod_{\substack{j=1 \\ j \neq n}}^N(1-P_{cs,j})$.

\subsection{Compatibility to the IEEE 802.15.4 Standard}

The superframe structure for the proposed models is similar to the standard IEEE 802.15.4 superframe structure~\cite{stand} as shown in Fig.~\ref{structure}. The IEEE 802.15.4 standard MAC can be considered to be a special case of the proposed models (i.e., $M = 7$ and each node is allowed to use a maximum of one slot in a superframe). The coordinator can assign an index number ($n = 1,2, \cdots,N$) to the associated node. In the MCCA scheme, the guaranteed time slot (GTS) list field of the beacon frame in the IEEE 802.15.4 can be modified to include the actions and assigned TDMA slot numbers in a format shown in Fig.~\ref{s_list}.

In the MDCA scheme, we assume that each packet contains two bits of overhead. The first bit is set if a TDMA slot request is sent and the second bit is set if a TDMA slot de-allocation request is sent. This modification removes the burden of sending separate packet for the TDMA slot request and de-allocation requests. Similarly, the acknowledgment packet consists of few bits ($\left\lceil \log_2(M+1)\right\rceil$ bits) of overhead  to notify the assigned TDMA slot number. After receiving a TDMA slot request, the coordinator allocates a TDMA slot to a node in a first-come first-served (FCFS) fashion. Note that in the standard, the guaranteed time slot (GTS) is used for time-critical data transmission. In our case, the purpose of using TDMA slots is to reduce network congestion during CAP. The proposed MDCA scheme would be compatible to the IEEE 802.15.4 standard MAC if the standard protocol is enhanced to decode the overhead bits in data packet as the GTS request and the GTS-deallocation request, and the overhead bits in acknowledgment packet as notification of TDMA slot allocation.

\section{MDP-Based Distributed Channel Access (MDCA) Model}

In this section, we want to determine which action is the best when a node has packet buffer level $b$ under the condition that the number of nodes in the network is $N$, the length of CAP  is $K-M$ slots, and the length of CFP is $M$. We call this as a policy $\pi_{N,K,M}$ of a node. We define a set of actions that a node takes in each superframe as $\Lambda = \{a_1, a_2, a_3, a_4\}$, where
\begin{itemize}
\item $a_1$: go to low power mode (no transmission)
\item $a_2$: transmit data packets during CAP
\item $a_3$: transmit data packets during CFP
\item $a_4$: transmit data packets during CAP and CFP.
\end{itemize}
Let us define the state of node $n$ at superframe $t$  as $S_{t,n} = B_{t,n}$, where $B_{t,n}$ is the buffer state. The buffer state $B_{t,n}$ is defined as the number of packets in the buffer of node $n$ at superframe $t$ such that $B_{t,n} \in \{0, 1, \dots, {B_{max}}\}$, where $B_{max}$ is the \textcolor{red}{maximum number of packets stored in the buffer for transmissions.} 
At each buffer state of node $n$ at superframe $t$, node $n$ takes one of the actions denoted as $A_{t,n} \in \Lambda$. To realize it, we assume that $M$ slots are randomly assigned to $N$ nodes. Note that if all nodes take action $a_2$, the CAP becomes congested while the CFP remains unoccupied. Similarly, if all nodes take action $a_3$, the CAP remains unoccupied whereas the TDMA slots in CFP become congested given $M < N$. To balance the use of CAP and CFP, we formulate the problem of decision making on packet transmissions during CAP or CFP or both, or no transmission at all by using an infinite-horizon Markov Decision Process (MDP).  An MDP is described by its  states, actions, reward, and transition probabilities.


For distributed channel access, a node assumes that other nodes also have packets to transmit and will compete to get access to the channel during CAP. Therefore, $\alpha$, $\beta$, $P_{c}$, and $\Phi_{cap}$ are estimated analytically for given $T_{cap}$ and $N$ in the saturated mode (i.e., a node assumes that all other nodes in the network have packets to send)~\cite{patro}.  For given $T_{cap}$, $N$, and packet arrival rate $\lambda$ at the saturation region by solving the infinite-horizon MDP problem we develop the transmission policy for a node.

In the MDCA scheme, we focus on the operation of one node. Therefore, we omit the node index $n$ from all the notations. For example, the buffer state is denoted by $B_t$ instead of $B_{t,n}$.




\subsection{Reward}

Let $R_{s,a}$ be the reward that a node receives for taking action $A_t = a$ at state $S_t = s$ at a superframe $t$. If a node defers the transmission, it saves energy but its buffer level may remain the same or increase. When the node transmits during both CAP and CFP, its throughput increases but it consumes a significant amount of energy. 
We define the expected reward for taking action $a$ at state $s$ as
\beq
R_{s,a} = \frac{\mu_{s,a} - s}{\max(s,1)} - \frac{\Xi_{s,a}}{\Xi_{max}} - C_{s,a}
\eeq
where $\mu_{s,a}$, $\Xi_{s,a}$, and $C_{s,a} \in [0,1]$ are the MAC throughput (number of packets retrieved out of the MAC buffer per superframe), energy consumed in joule, and bandwidth cost, respectively, for taking action $a$ at state $s$ and $\Xi_{max}$ is the maximum energy consumed. 
\textcolor{red}{ A node with a higher buffer state $(s)$ needs to transmit more packets. However, transmission of more packets results in a higher energy consumption. The physical meaning of the reward function is that a node tries to improve the throughput while reducing energy consumption in the unwanted transmissions. The bandwidth cost penalizes those nodes which try to occupy the time slots even though they do not have any packets to transmit.}

Let $\Xi_{x}$ denote the energy (in Joule) required to transmit a packet and let $\Xi_{c}$ denote the energy (in Joule) required to perform carrier sensing. The total amount of energy required to transmit a packet during the CAP is \textcolor{red}{obtained based on the probability of successful packet transmission and probability of going to backoff stages before a packet is successfully transmitted as follows:}
\beq
\Xi_p = \frac{1 - P_c^{W+1}}{1 - P_c}\Xi_{x} + \frac{1 - P_c^{W+1}}{1 - P_c}\frac{1 - \phi^{m+1}}{1 - \phi} \Xi_{c}
\eeq
where $\phi = (1 - \alpha \beta)(1-P_d)$ is the probability of going to another backoff stage with $P_d$ being the probability of transmission being deferred, $m$ is the maximum number of backoffs allowed, and $W$ is the number of retransmissions allowed. The amount of energy consumed for taking action $a$ at state $s$ is
\beqn
  \Xi_{s,a} &=& \begin{cases} \min(\kappa,s) \Xi_p,  \mbox{if} \ \  a = a_2  \nn \\
                       \min(\eta,s) \times (2\Xi_{x}),  \mbox{if} \ \  a = a_3 \nn \\
                       \min(\kappa,\max(s-\eta,0)) \Xi_p + \min(\eta,s) \times (2\Xi_{x}), \;  \mbox{if} \ \  a = a_4  \nn \\
                          0,  \mbox{otherwise} \ \  \end{cases}
\eeqn
where $\kappa$ is the goodput expressed in number of packets per superframe. The MAC throughput depends on action $a$ taken at state s and is expressed as
\beqn
\mu_{s,a} &=& \begin{cases} \min(\Phi_{cap},s),  \mbox{if} \ \  a = a_2  \\
                       \min(\eta,s),  \mbox{if} \ \  a = a_3 \nn \\
                       \min(\Phi_{cap},\max(s-\eta,0)) + \min(\eta,s),  \mbox{if} \ \  a = a_4 \nn \\
                          0,  \mbox{otherwise}. \ \  \end{cases} \nn
\eeqn

In the above equation, for the purpose of calculation of relative energy, we set $\Xi_{max}= s\Xi_p$.
\textcolor{red}{The nodes which have no packet to transmit but tries to occupy the time slot are penalized the most, (i.e., $C_{s,a} = 1$ for $s = 0$). The nodes which have some packets are penalized lightly, otherwise they are not penalized (i.e., $C_{s,a} = 0$).}
 \begin{numcases}{\label{eqnreward} C_{s,a} =}
1 - \frac{s}{\eta}, & \mbox{if}  \ \  $s \leq \eta$ \mbox{ and }  $a \in \{a_3, a_4\}$  \nn \\
0, &  \mbox{otherwise.} \nn
\end{numcases}

\subsection{State Transition Probability}

When a node is in state $s=b$ during superframe $t$, the probability of going to state $s'=b'$, when action $a$ is taken, is given by
\beq
\Pr[S_{t+1} = s'| S_t = s, A_t=a] = \Pr[B_{t+1}=b'|B_{t} = b, A_t=a].
\eeq

When action $a$ is taken, the probability that the buffer state changes from $b$ to $b'$ is given by the probability of arrival of $x = \left\lceil b' - b + \mu_{s,a}\right\rceil$ packets at the beginning of superframe $t+1$, i.e., the buffer state transition probability
\beqn \label{buff1}
  \Pr[B_{t+1}=b'|B_{t}=b, A_t = a] &=& \Pr\mbox{[arrival of} \ x \  \mbox{packets]} \nonumber  \\
                        &=& \begin{cases} f^X_{\Gamma}(x), & \mbox{if} \ \  x \geq 0 \nn \\
                        0, & \mbox{otherwise} \end{cases}
\eeqn
where $\Gamma = T_{sf} + T_{beacon}$ and $f^X_{\Gamma}$ is the probability mass function of number of packet arrivals $X_{\Gamma}$. Similarly, when the next buffer state is $B_{max}$, $x = \left\lceil B_{max} - b + \mu_{s,a}\right\rceil$, and
\beqn \label{buff2}
\Pr[B_{t+1} = B_{max}|B_{t}=b,A_t = a] =  & \nn\\
\Pr\mbox{[number of packet arrivals} \geq x \mbox{]}   & 
 = 1 - \sum_{h = 0}^{x-1} f^X_{\Gamma}(h).  \nn
\eeqn

\subsection{MDP Solution}

Let $\pi^*_s$ be the policy that maps a state $s$ into an action $a$ and $V$ be the value function corresponding to the total expected discounted reward over an infinite horizon. The objective is to maximize the total expected reward. The optimal value function $V^*$ is expressed by the Bellman optimality equation~\cite{puterman} as follows:
\beq \label{bell}
 V^{*}_s = \max_{\substack{a \in \Lambda}}\left(R_{s,a} + \gamma \sum_{s' \in \Upsilon} \Pr[S_{t+1} = s'|S_{t} = s, A_t = a] V^{*}_{s'}\right)
\eeq
for all $s \in \Upsilon $, where $\Upsilon$ is the set of all possible states, $R_{s,a}$ is the expected value of the reward, and $\gamma \in [0,1)$ is the discount rate. The Bellman equation can be solved by the value iteration method to find $V^{*}~\cite{puterman}$.
The optimal policy $\pi^*_{s}$ for all $s \in \Upsilon $, is given by, $\pi^*_{s} =$
\beq
\arg \max_{\substack{a \in \Lambda}}\left(R_{s,a} + \gamma \sum_{s' \in \Upsilon} \Pr[S_{t+1} =s'|S_{t} =s, A_t = a] V^{*}_{s'} \right). \nn
\eeq

The value iteration method requires $(\left| \Lambda\right|\left|\Upsilon\right|^2)$ computations per iteration~\cite{littman}. Note that the policy iteration method requires a fewer number of iterations to find the optimal policy. However, it requires more computations per iteration than the value iteration method. As described in~\cite{puterman}, the value iteration method converges to the optimal solution in a finite number of iterations at a rate of $\gamma$ if the stopping criterion is $\epsilon \frac{(1-\gamma)}{2\gamma}$ for $\epsilon > 0$. 



\section{MDP-Based Centralized Channel Access (MCCA) Model}

\subsection{MDP Formulation}

With the MDCA method, the nodes are unaware of the actions of the other nodes. This suggests that the method can be improved by using a centralized approach. In this section, we present a method in which the coordinator determines the policy based on the buffer status of all the nodes.

We assume that the coordinator has the knowledge of the distribution of packet arrival  at all the nodes. In this method, the buffer level represents the state of a node. The state of the network is defined as $\mathbf{S}_t = \mathbf{B}_t$, where $\mathbf{B}_t = (B_{t,1},B_{t,2}, \cdots,B_{t,N})$ denotes the joint buffer state of $N$ nodes during a superframe $t$ and $B_{t,n} \in \{0,1,\cdots,B_{max}\}$ is the buffer state for node $n$. Let $\mathbf{A}_t = (A_{t,1},A_{t,2}, \cdots, A_{t,N})$ denote the joint actions of $N$ nodes, where $A_{t,n} \in \{a_1, a_2, a_3, a_4 \}$.
Given any state  $\mathbf{b} = (b_1,\cdots,b_N)$ and action $\mathbf{a} = (\bar{a}_1,\cdots,\bar{a}_N)$, let $\mathbf{b}' = (b'_1,\cdots,b'_N)$ denote the next joint state. We define the joint reward as $\mathbf{R}_{\mathbf{b},\mathbf{a}} = \sum_{n=1}^N R_{b_n,\mathbf{a}}$, where $R_{{b_n},\mathbf{a}}$ is the reward of node $n$. Similar to the MDCA scheme, the reward is given by
\beq
R_{b_n,\mathbf{a}} = \frac{\mu_{b_n,\mathbf{a}} - b_n}{\lambda} - \frac{\Xi_{b_n,\mathbf{a}}}{\Xi_m}
\eeq
where $\lambda$ is the average number of packet arrivals per superframe duration, $\Xi_m = \Xi_x \eta T_{cap}/T_{slot}$ and $\mu_{b_n,\mathbf{a}}$ is the MAC throughput of node $n$ when the joint action by all the nodes in the network is $\mathbf{a}$. The transition probability is defined as $\Pr[\mathbf{S}_{t+1} = \mathbf{b}'| \mathbf{S}_t = \mathbf{b}, \mathbf{A}_t= \mathbf{a}] = \prod_{n=1}^N \Pr[B_{t+1,n} = b'_i|B_{t,n} = b_n,\mathbf{A}_t = \mathbf{a}] $.  Similar to~(\ref{buff1}) and~(\ref{buff2}), the probability that a node $n$ goes to buffer state $b_{n}'$ from state $b_{n}$ is given by
\beq
   \Pr[B_{t+1,n} = b_{n}'|B_{t,n} = b_{n},\mathbf{A}_t= \mathbf{a}] = \Pr\mbox{[arrival of} \ x_n \  \mbox{packets]} \nn
\eeq
where $x_n = b_{n}' - b_{n} + \mu_{b_n,\mathbf{a}} $.

The coordinator solves the MDP problem and determines the optimal policy for each state. During a superframe, the coordinator observes the buffer level of all nodes  to determine the state and broadcasts the optimal policy. A node can piggyback the information of buffer level to the coordinator while transmitting data packets. However, at the coordinator, the information about the buffer level at a node would not be accurate when the node is unable to transmit any packet successfully during a superframe and the packet arrival rate at that node is not deterministic. In particular, when packet arrival takes place at the beginning of the superframe, the piggybacked information of the buffer level would be inaccurate. For this reason, the coordinator has to take into account the  time of receiving the buffer level information and the average number of packet arrivals during a superframe period.

For each node, the coordinator has to keep the latest buffer level report as well as the index of the frame in which the latest buffer level report was received. In frame $t$, the coordinator maintains the buffer level information for node $n$ in the form of the tuple ${\mathbf G}_{t,n}=(Q_{t,n},F_{t,n})$, where $F_{t,n}$ is the number of superframes which have passed after the latest report was received and $Q_{t,n}$ is the buffer level in the latest report of the node $n$. The coordinator estimates the average buffer level of node $n$ as
\beq \label{buf_level}
 \bar{Q}_{t,n} = Q_{t,n} + \left\lfloor \lambda_n F_{t,n}\right\rfloor.
\eeq
The buffer state of a node $n$ is determined as $B_{t,n} = \bar{Q}_{t,n}$ if $\bar{Q}_{t,n} < B_{max}$, otherwise $B_{t,n} = B_{max}$.

\subsection{Complexity of Solving the MDP Problem}
The coordinator finds the optimal policy $\pi^*_{\mathbf{S}}$ for any state $\mathbf{S}$ by solving the Markov decision problem. The coordinator sends the policy information to the nodes through the beacon frame. However, the complexity is huge because of the large dimensions of state and action. For a network of size $N$, the value iteration method has a computational complexity of $\mathcal{O}(4^N (B_{max}+1)^N)$.
Therefore, finding an optimal solution is not practical. We propose an approximate solution in the next section.

\subsection{Approximate Solution}
This solution (the procedure of which is described in \textbf{Algorithm~\ref{algocent}}) is based on the assumption that nodes with higher buffer occupancy level are unlikely to defer their transmissions and are highly likely to use a TDMA slot during CFP. In the literature, the longest-queue-first (LQF) scheduling scheme during CFP has been shown to be throughput maximal~\cite{Lecon}. Also, instead of letting all the nodes to compete during CAP, some nodes can be put into  the
low-power mode so that congestion is reduced during CAP and throughput is improved. In this section, we present a solution which combines the merits of the LQF scheduling scheme and a congestion reduction scheme. In the latter scheme,  $N' \leq N$ nodes are allowed to  transmit  during the CAP such that  for given system parameters the saturation throughput is maximized. If $M$ nodes are allocated TDMA slots, then  the remaining $N-N'-M$ nodes with relatively lower buffer occupancy levels are put into low power mode (or no transmission mode).

\begin{algorithm}[H]
 \caption{Approximate solution for centralized MDP}\label{algocent}
 \begin{algorithmic}[1]
 \STATE Input: Buffer level at all the nodes $\mathbf{q} = (q_{1},q_{2},\cdots,q_{N})$, number of slots in CFP $M$
 \STATE Output: $\mathbf{a}$
 \STATE Sort nodes $\mathbf{d} = \left\{1,2, \cdots, N \right\}$ such that $q_{n} \geq q_{n+1}, \ \  \forall n \in \mathbf{d}$
 \STATE for each element $\mathbf{d}_{g} \in \{ \emptyset,\{1\},\{1,2\}, \cdots,\{1,2,\cdots, M\} \}$ do
 \STATE \ \ for each element $\mathbf{d}_{dg} \in \{ \emptyset,\{j\},\{j,j+1\}, \cdots,\{j,j+1,\cdots, M\} \}$, for $j = \left|\mathbf{d}_{g}\right| + 1$ do
 \STATE \ \ for each element $\mathbf{d}_{d} \in \{ \{j,j+1\},\{j,j+1,j+2\}, \cdots,\{j,j+1,j+2, \cdots, N\} \}$, for $j = \left|\mathbf{d}_{g}\right| + 1$ do
 \STATE \quad Calculate utility $\mathbf{u}_{\mathbf{d}_{g},\mathbf{d}_{dg},\mathbf{d}_{d},\mathbf{d}_{s}} = \sum_{n=1}^N (\frac{\mu_{n} - q_n}{\lambda} - \frac{\Xi_{n}}{\Xi_{max}})$ where
 \STATE \quad $\mu_{n} = \min(q_{n},\eta)$ and $\Xi_{n} = \mu_n \Xi_{x}$ for $n \in \mathbf{d}_{g}$
 \STATE \quad $\mu_{n} = \min(q_n,\eta) + \min(\Phi_{cap}, \max(0,q_n-\eta)  $ and \\ \quad $\Xi_{n} = \min(q_n,\eta) \Xi_{x} + \min(\kappa, \max(0,q_n-\eta)\Xi_{p}$ for $n \in \mathbf{d}_{dg}$
 \STATE \quad $\mu_{n} = \min(\Phi_{cap},q_n)$ and $\Xi_{n} = \min(\kappa,q_n) \Xi_{p}$ for $n \in \mathbf{d}_{d}$, $n \notin \mathbf{d}_{dg}$
 \STATE \quad $\mu_{n} = 0$, $\Xi_{n} = 0$, $\mathbf{d}_{s} \leftarrow n$ otherwise \\
  \quad where throughput $\Phi_{cap}, \kappa$ are calculated for given $|\mathbf{d}_{d}|$ and $M$
 \STATE \ \  end for
 \STATE \ \ end for
 \STATE end for
 \STATE Find $\mathbf{a} = \left\{\mathbf{d}_{g},\mathbf{d}_{dg},\mathbf{d}_d,\mathbf{d}_{s}\right\}$ for $\max \mathbf{u}_{\mathbf{d}_{g},\mathbf{d}_{dg},\mathbf{d}_{d},\mathbf{d}_{s}} $

 \end{algorithmic}
\end{algorithm}

The coordinator observes the buffer level $\bar{Q}(t,n)$ of the nodes $n \in N$ at the beginning of the superframe $t$. Note that from~(\ref{buf_level}), $\bar{Q}(t,n)$ might be higher than $B_{max}$. It sorts the nodes in the descending order of their buffer levels. It calculates the utility function (defined in step 7 in~\textbf{Algorithm~\ref{algocent}}) for every combination of the actions provided that only the first $M$ nodes are allowed to use the TDMA slots. The utility function is the same as the reward function presented earlier. The coordinator determines the set of best actions of all the nodes  $\mathbf{a}$ that gives the maximum value of the utility function and sends it through the beacon frame.  It can memorize the best action vector $\mathbf{a}$ for the given state $\mathbf{S}$ to use it next time. In the algorithm, $\mathbf{d}_s, \mathbf{d}_d, \mathbf{d}_g$, and $\mathbf{d}_{dg}$ are the sets of nodes taking the actions $a_1$, $a_2$, $a_3$, and $a_4$, respectively.

Let $\mathbf{\mathcal{A}}$ be the set of all possible action vectors and $\mathbf{U}$ be the set of all possible utility functions. \textbf{Algorithm 1} has a computational complexity of $\mathcal{O}(N\mbox{log}N + \left|\mathbf{\mathcal{A}}\right|)$, where $\left|\mathbf{\mathcal{A}}\right|$ depends on the number of utility functions to be computed at a state $\mathbf{S}$, and is given by
\beqn
  \left|\mathbf{\mathcal{A}}\right| &=& \left|\mathbf{U}\right| = \sum_{n = 1}^{\left|\mathbf{D}_{g}\right|} \sum_{j = 1}^{\left|\mathbf{D}_{dg}\right| + 1 - n} \sum_{h = 2}^{N+1-n} 1\nn \\
                          &=& \sum_{n = 1}^{\left|\mathbf{D}_{g}\right|} (\left|\mathbf{D}_{dg}\right| + 1 - n)(N-n)
\eeqn
where $\mathbf{D}_{g}$ and $\mathbf{D}_{dg}$ are the sets of all possible elements $\mathbf{d}_{g}$ and $\mathbf{d}_{dg}$, respectively. Suppose $M = 7$, then $\left|\mathbf{D}_{g}\right| = \left|\mathbf{D}_{dg}\right| = 8$ and $\left|\mathbf{\mathcal{A}}\right| = 36N - 120$.  The coordinator determines the policy for the nodes at the beginning of each superframe.
\section{Extension of the Models Considering  Channel Fading}

In this section, we present a methodology for the calculation of the parameters ($\alpha_n, \beta_n, P_{c,n} \ \forall n \in \{1, 2, \cdots, N \}$) considering channel fading. The key idea to extend the MDP-based models presented earlier by considering the presence of channel fading is to determine the correct parameters and the throughput ($\Phi_{cap}$). It is assumed that channel fading remains the same during packet transmission time.

Due to signal attenuation in the channel, the transmission range is reduced and so is the carrier-sensing range. Due to the reduced transmission range, the network suffers from outage as well as hidden node collision problem. When the received signal level falls below the receiver threshold, the transmission suffers outage because the receiver cannot decode the signal successfully. For short-range networks such as personal-area networks~\cite{bharattitb}, signal attenuation can be modeled by using distance-dependent attenuation along with log-normal shadowing.
If $\Omega_{tx}$ is the transmit power in dB, $\ell(\nu_n)$ is the loss (in dB) for transmission from a node $n$ to the coordinator with separation of $\nu_n$, and $\zeta$ is the shadowing component with zero mean and standard deviation of $\sigma$ (e.g., 4.4 dB)~\cite{yazdan}, then the received power (in dB) is:
$\Omega_{rx} = \Omega_{tx} - \ell(\nu_n) - \zeta.$
The  probability that the received power is less than the receiver threshold $\psi$ dB (i.e., outage probability) is given by
\beqn
\Theta_n =\Pr[\Omega_{rx} < \psi] = 1 - \frac{1}{2} \mbox{erfc}\left(-\frac{\Omega_{tx} -\ell(\nu_n) - \psi}{\sqrt{2}\sigma}\right)
\eeqn
where $\mbox{erfc()}$ is the complementary error function. An example of the propagation model for signal attenuation~\cite{bharattitb} that can be considered is $\ell(\nu_n) = 27.6 \log(\nu_n[mm]) + 46.5 \log(2400[\mbox{MHz}]) - 157$.

Let $\nu_{nj}$ be the distance between node $n$ and node $j$ and $\xi$ be the carrier-sensing threshold in dB. The channel fading gains of the links $n,j \in N$ are independent. Even though there is no outage in the link between a node and the coordinator, it is probable that the node is hidden to other nodes transmitting in the different links. Then, the probability that node $n$ and node $j$ are hidden to each other is
\beq
H_{n,j} = 1 - \frac{1}{2} \mbox{erfc}\left(-\frac{\Omega_{tx} -\ell(\nu_{nj}) - \xi}{\sqrt{2}\sigma}\right). \nn
\eeq

Let $\Psi_n$ be the set of $\left| \Psi_n \right|$ nodes  which are hidden to node $n$ such that $H_{n, j} \neq 0, \  \forall j \in \Psi_n$. When a node $n$ transmits during CAP, the hidden node collision probability is estimated by the probability of channel being busy during first carrier sensing $P_{b,\Psi_n \cup \{n\}} = (1-\alpha_{n/ \Psi_n \cup \{n\}})$ when at least one node from $\Psi_n$ is transmitting among the nodes in the set $\Psi_n \cup \{n\}$~\cite{park}. We denote by $\alpha_{n/ \Psi_n \cup \{n\}}$ the probability of channel being idle in the first carrier sensing for node $n$ given the nodes in the set $\Psi_n \cup \{n\}$. The hidden node collision probability for node $n$ is estimated as
\beqn
 H_{n} &=& \sum_{j=1}^{\left|\Psi_n\right|_1} H_{n,j} \prod_{^{h=1}_{h \neq j}}^{\left| \Psi_n\right|}(1-H_{n,h}) P_{b,\{n,j\}} + \\
            &   & \sum_{j=1}^{\left|\Psi_n\right|_1} \sum_{^{r=1}_{r\neq j}}^{\left|\Psi_n\right|_2} H_{n,j}H_{n,r} \prod_{^{h=1}_{h \neq j,r} }^{\left|\Psi_n\right|} (1-H_{n,h}) P_{b,\{n,j,r\}} + \nonumber\\
        & & \cdots + \sum_{j=1}^{\left|\Psi_n\right|_1} \sum_{^{r=1}_{r\neq j}}^{\left| \Psi_i\right|_2} \cdots \sum_{^{l=1}_{l\neq jr} }^{\left|\Psi_n\right|_{\left| \Psi_n\right|}}  H_{n,j}H_{n,r} \cdots H_{n,l} \nn \\
         & & \prod_{^{h=1}_{h \neq j,\cdots,l}}^{\left| \Psi_i\right|}(1-H_{n,h}) P_{b,\{n,j,r, \cdots, l\}}.  \label{hi}
\eeqn
For example, if $\Psi_1 = \{ 3,4 \}$, then $H_{1} = H_{1,3}(1-H_{1,4})P_{b,13} + H_{1,4}(1-H_{1,3})P_{b,14} + H_{1,3}H_{1,4}P_{b,134}$. As in (\ref{pctilda}), when both the channel fading and hidden node collision are taken into account, the probability of error is calculated as $\hat{P}_{c,n} = \left(P_{c,n}(1-H_n) + H_n\right)(1-\Theta_n) + \Theta_n.$
In a similar way, we derive $\alpha_{n/\mathbf{d}}$, where $\mathbf{d} = \{1, 2, \cdots, N\}$ is the set of $N$ nodes, as follows:
\beqn
 \alpha_{n/\mathbf{d}} &=& \prod_{h=1}^{\left|\Psi_i\right|}(1-H_{n,h}) \alpha_{n/\mathbf{d}} +\sum_{j=1}^{\left|\Psi_n\right|_1} H_{n,j}\prod_{^{h=1}_{h \neq j} }^{\left|\Psi_n\right|}(1-H_{n,h}) \alpha_{n/\mathbf{d} \setminus\{j\}} \nn \\
      & & + \sum_{j=1}^{\left| \Psi_n\right|_1} \sum_{^{r=1}_{r\neq j} }^{\left|\Psi_n\right|_2} H_{n,j}H_{n,r}\prod_{^{h=1}_{h \neq j,r} }^{\left|\Psi_n\right|}(1-H_{n,h}) \alpha_{n/\mathbf{d} \setminus \{j\} \cup \{r\}} \nn \\
        & & + \cdots  + \sum_{j=1}^{\left|\Psi_n\right|_1} \sum_{^{r=1}_{r\neq j}}^{\left|\Psi_n\right|_2} \cdots \sum_{^{l=1}_{l\neq j,r} }^{\left|\Psi_n\right|_{\left| \Psi_n\right|}}  H_{n,j}H_{n,r} \cdots H_{n,l} \nn \\
         & & \prod_{^{h=1}_{h \neq j,\cdots,l}}^{\left|\Psi_i\right|}(1-H_{n,h})\alpha_{n/\mathbf{d} \setminus \{j\} \cup \{r\} \cdots \cup \{l\}}.
\eeqn
Similarly, we derive $\beta_{n/\mathbf{d}}$ for node $n$. Even though the nodes are homogeneous, their positions lead to  heterogeneity in the network. Given $\mathbf{P}_{cs} = \{P_{cs,1}, \cdots, P_{cs,N}\}$, we calculate and update  $P_{c,n}, \alpha_n, \beta_n, \ \forall n \in \mathbf{d}$ by solving the Markov chain model (see~\cite{bharattitb} and~\cite{park} for the details of the Markov chain model) until $\left|\mathbf{P}_{cs}^{c+1} - \mathbf{P}_{cs}^c\right| < \delta $ after $c$ iterations, $\delta$ is a small positive number. After determining the new parameters considering channel fading, we calculate the CAP throughput $\Phi_{cap,n}$. We also calculate the CFP throughput as $(1-\Theta_n)\min(\eta,b_n)$, where $b_n$ is the buffer level of node $n$.

However, if each node is considered to be within the carrier-sensing range of the other nodes when the statistical variation in the channel propagation condition is not considered, carrier sensing range is at least double the transmission range and all the links between nodes in the network suffer same channel fading, there will be no effect of hidden node collision  on the packet reception at the coordinator. This is because, when the probability of channel outage is zero,  the hidden node collision probability also becomes zero. Hidden node collision will occur  when there is outage at the coordinator. In this case, the probability of channel outage is sufficient to update the collision probability, i.e., $\hat{P}_{c,n} = \left(P_{c,n}(1-\Theta_n) + \Theta_n\right)$.

\section{Performance Evaluation}

\subsection{Performance Metrics and Simulation Parameters}

For performance evaluation, we simulate the proposed channel access schemes in MATLAB. We consider packet delivery ratio (PDR), end-to-end delay, and energy consumption rate as the performance metrics. The packet delivery ratio (PDR) is defined as the ratio of the number of packets successfully transmitted and number of packets generated by the nodes during the simulation run time. The end-to-end delay is measured from the time a packet is generated until it is successfully transmitted. The average energy consumed by nodes (including the coordinator) per successfully transmitted packet in the network is considered as the energy consumption rate metric. For performance evaluation, we use the power consumption values for an IEEE 802.15.4 transceiver as follows~\cite{chipcon}:  power consumption in sleep mode,  transmit mode, receive mode, and idle mode is 36 $\mu$W, 31.32 mW, 33.84 mW, and 766.8 $\mu$W, respectively.


We consider a star network topology consisting of a coordinator and $N = 20$ nodes placed in a circle with a transmission range of 10m and a carrier sensing range of 20m. Each node is within the carrier-sensing range of other nodes when channel fading is not considered. Each node transmits packets to the coordinator located at the center. We assume that the hybrid MAC protocol operates with a physical data rate of 250 Kbps. The smallest unit of time, i.e., the unit backoff period (UBP), is $320 \mu s$. Unless otherwise specified, we assume that there is no packet loss due to channel fading. We set the discount factor $\gamma$ to 0.9. For performance evaluation, we set $\Xi_{x}$ to 1 J and $\Xi_{c}$ to 0.1 J.
We consider the buffer size of $B_{max} = 5$. 
We assume a fixed batch size of length one. \textcolor{red}{We assume $P_{discard} = 0$ and $P_{drop} = 0$ when calculating the MAC layer throughput.}

Unless otherwise specified, we consider the physical packet size of $6$ UBP (i.e., 60 bytes long), the acknowledgment packet size of 1 UBP and inter-frame space of 1 UBP. Since a packet has to be transmitted at the boundary of the UBP, a successful packet transmission time including propagation time, inter frame space (IFS) and acknowledgment would be $T_{tx} = 10$ UBP. We also consider the beacon frame length to be 4 UBP. We assume that a node can transmit $\eta = 2$ packets per slot duration. To achieve this, we set $K = 16$, $M = 7$ (similar to the superframe structure of the IEEE 802.15.4 MAC standard). The superframe duration is $T_{sf} = 384$ UBP and the length of a slot is $T_{slot} = 24$ UBP. To prevent starvation of nodes from accessing the TDMA slots, we set $\varrho = 18$. In the figures, we define the offered traffic as $\frac{N\lambda T_{tx}}{T_{sf}+T_{beacon}}$. We run the simulations for 5000 beacon intervals.

\subsection{Simulation Results}

In this section, we present the performance evaluation results for the MDCA and MCCA schemes.  The superframe is divided into $K = 16$ slots. For slotted CSMA/CA, during contention period ($K-M$ slots), the set of contention window size is $cw \in [8,16,32,32,\cdots]$. Also, the nodes do not drop packets due to limits on the maximum number of backoffs  and retransmissions allowed. Note that acknowledgments are also required for the packets that are transmitted during the allocated TDMA slots.

\subsubsection{Comparison}
In the MDCA scheme, the transmission policy is  completely distributed (i.e., not determined by the coordinator). Therefore, we compare the MDCA scheme with the  slotted CSMA/CA scheme with default parameters of the IEEE 802.15.4 MAC in beacon-enabled mode with no CFP (i.e., $M$ = 0) and the contention control scheme (CCS) proposed  in~\cite{francesco}. The assumed MAC parameters for CSMA/CA are: $MACMaxBE = 5$, $MACMinBE = 3$, backoff limit $m = 4$, limit on the number of retransmissions $W = 3$. The contention control scheme (CCS) proposed in~\cite{francesco} tunes the protocol parameters such as contention window based on the required delivery ratio. The parameters for CCS are taken from Table I in~\cite{francesco}. Note that MDCA is an improved version of our previous work~\cite{bharatglb} which is hard to realize because each node requires high computational effort to solve the MDP problem. On the other hand, in the proposed MDCA scheme the MDP problem can be solved offline. For this reason, we do not include the scheme proposed in~\cite{bharatglb} in the comparison.

The MCCA scheme is compared with an existing centralized scheme called the adaptive CSMA/TDMA hybrid channel access (AHCA) scheme~\cite{Gilani}. The AHCA scheme is similar to LQF scheme and queue length-aware CSMA/TDMA hybrid channel access (QLHCA) scheme proposed in~\cite{Zhuo} under the system model of the proposed scheme.



\begin{figure*}[t]
\begin{minipage}[b]{0.3\linewidth}
\begin{center}
\includegraphics[width = 2.5in]{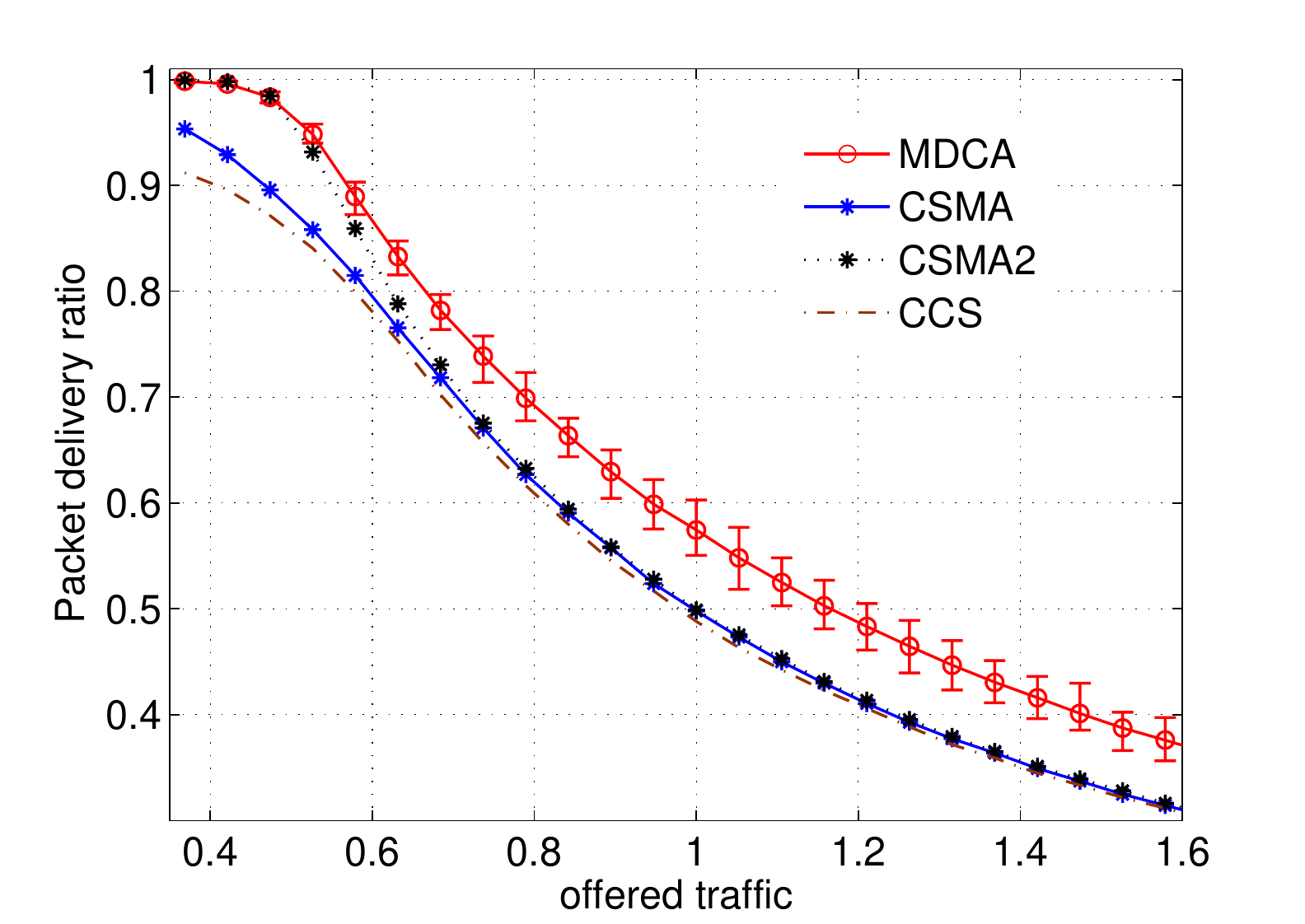}
\caption{Packet delivery ratio for different distributed schemes (for $N = 20$, $M = 7$,  $\eta = 2$). The error bar shows maximum and minimum values.}
\label{pdrSO3dist}
\end{center}
\end{minipage}
\hspace{0.2cm}
\begin{minipage}[b]{0.3\linewidth}
\begin{center}
\includegraphics[width = 2.5in]{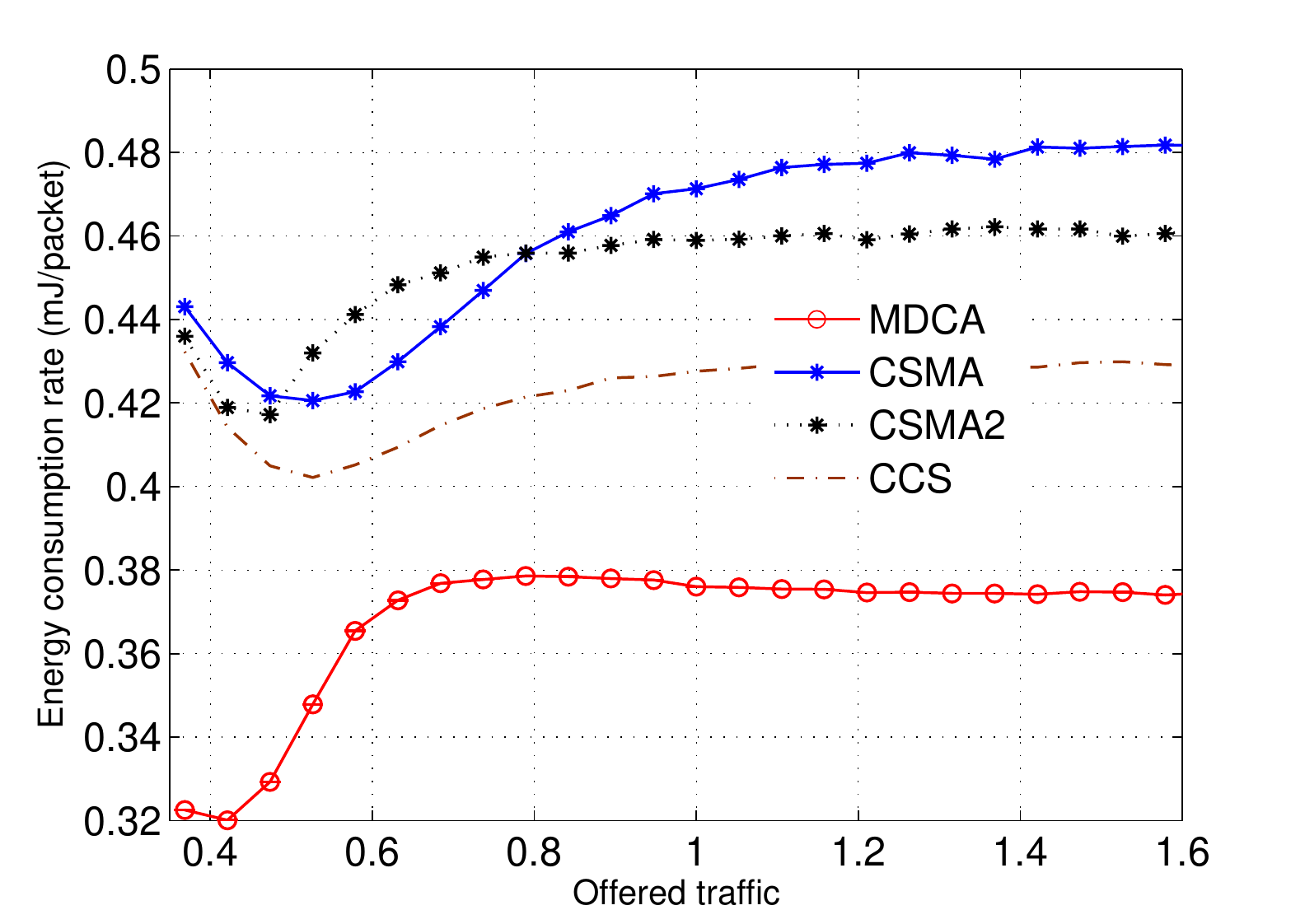}
\caption{Energy consumption rate for different schemes (for $N = 20$, $M = 7$, $\eta = 2$).}
\label{enSO3dist}
\end{center}
\end{minipage}
\hspace{0.2cm}
\begin{minipage}[b]{0.3\linewidth}
\begin{center}
\includegraphics[width = 2.5in]{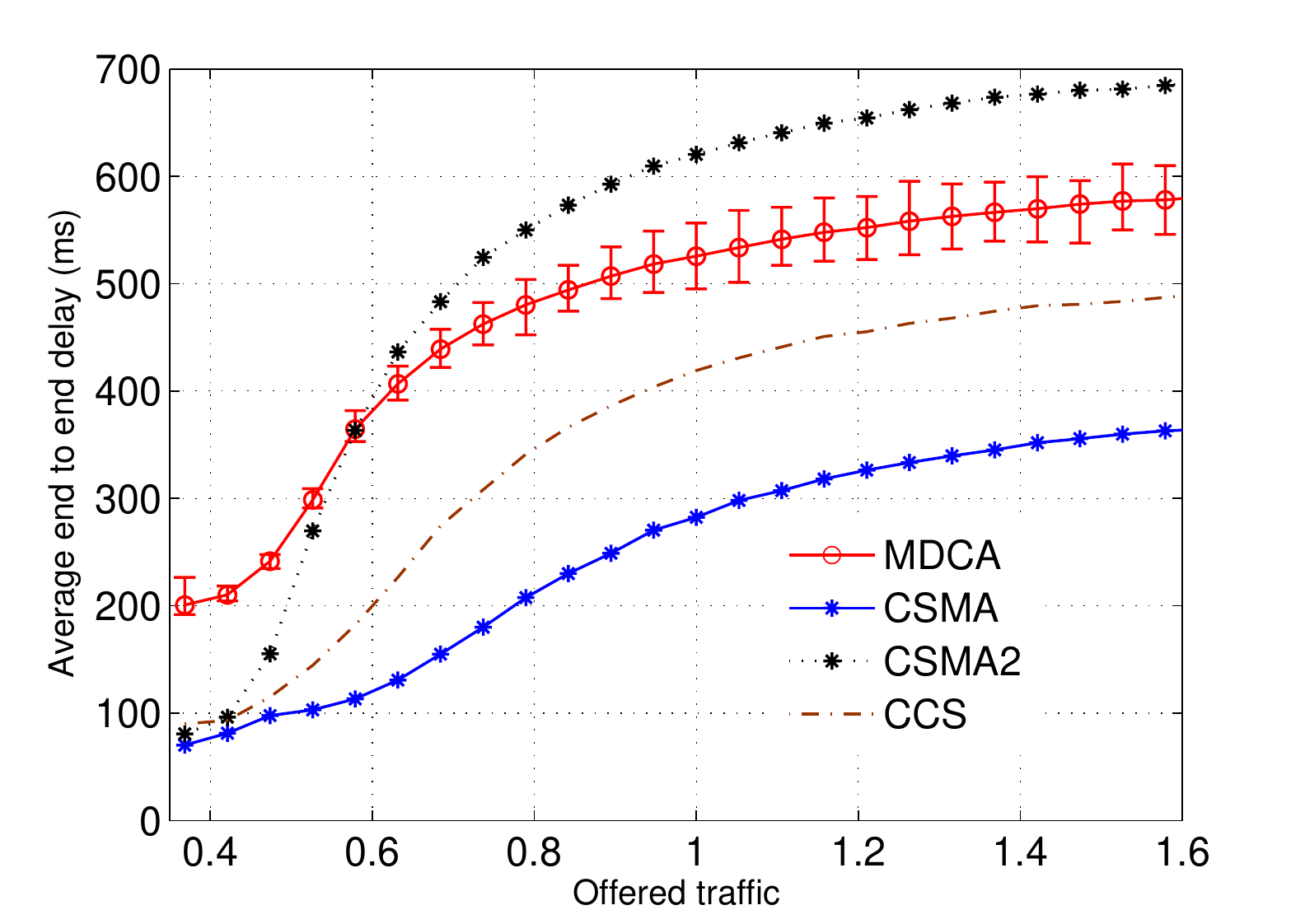}
\caption{Average end-to-end delay for different schemes (for $N = 20$, $M = 7$,  $\eta = 2$).}
\label{delSO3dist}
\end{center}
\end{minipage}
\end{figure*}

\begin{figure*}[t]
\begin{minipage}[b]{0.3\linewidth}
\begin{center}
\includegraphics[width = 2.5in]{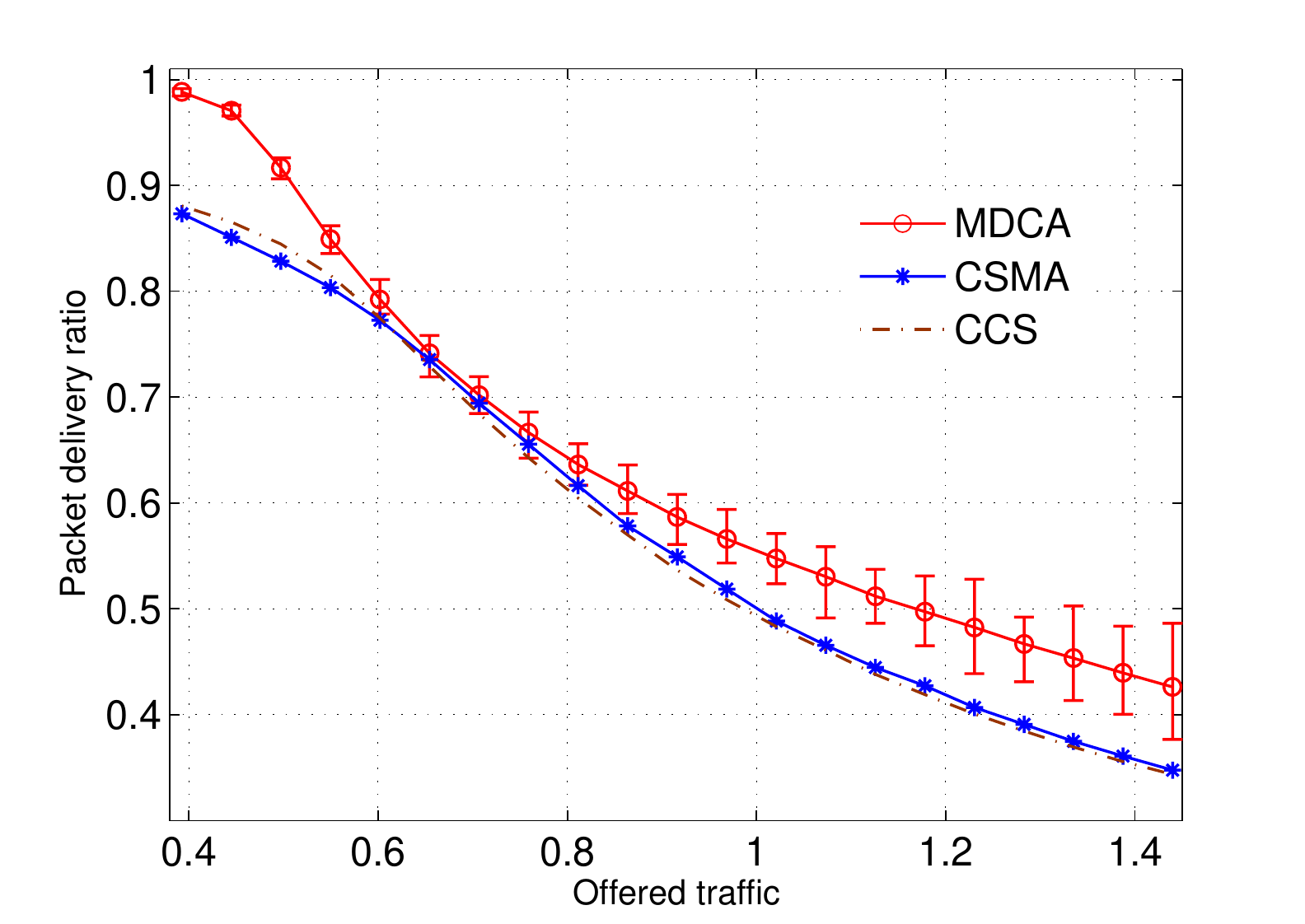}
\caption{Packet delivery ratio for different schemes (for $N = 20$, $M = 7$, $\eta = 4$).}
\label{pdrSO4dist}
\end{center}
\end{minipage}
\hspace{0.2cm}
\begin{minipage}[b]{0.3\linewidth}
\begin{center}
\includegraphics[width = 2.5in]{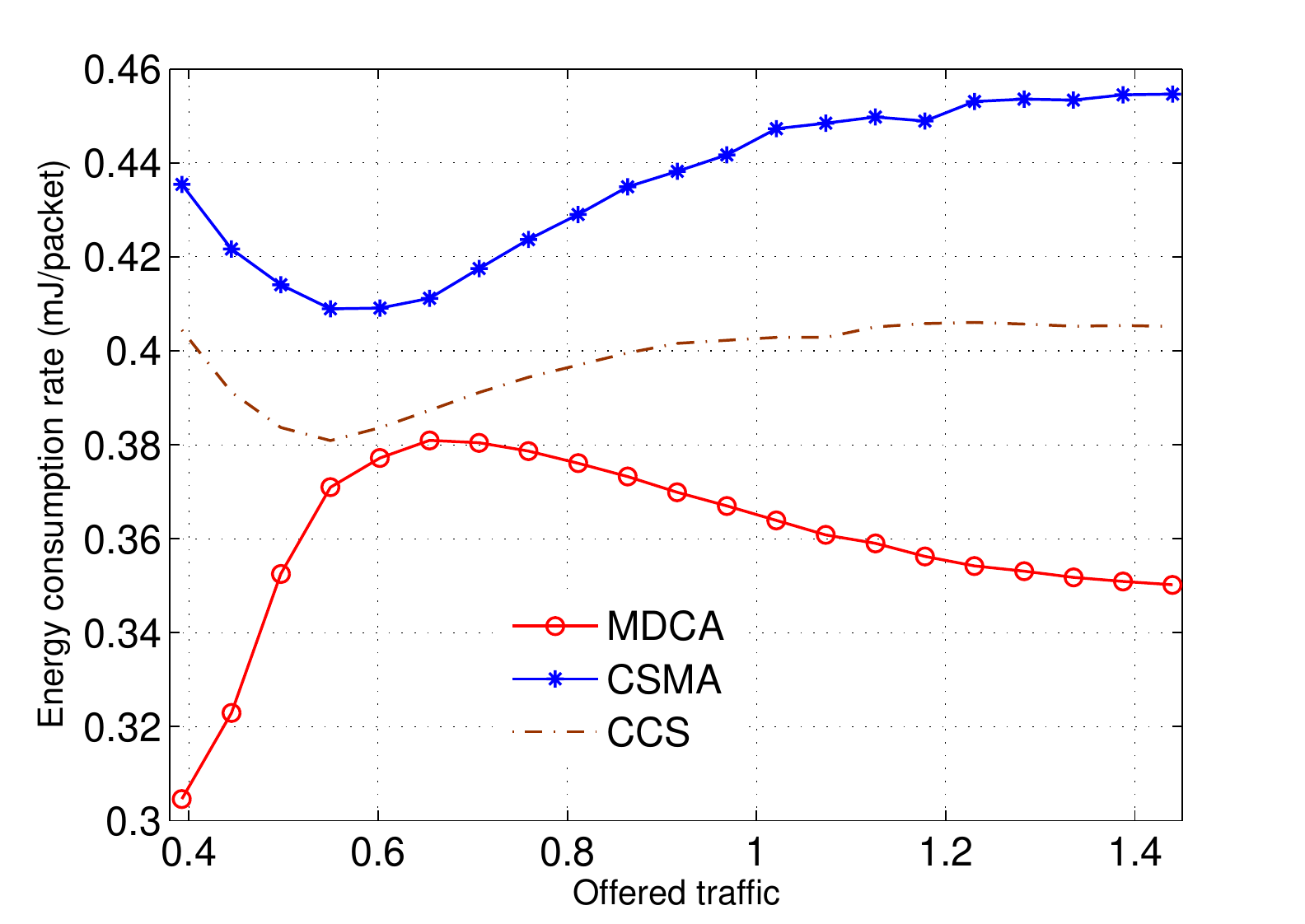}
\caption{Energy consumption rate for different schemes (for $N = 20$, $M = 7$,  $\eta = 4$).}
\label{enSO4dist}
\end{center}
\end{minipage}
\hspace{0.2cm}
\begin{minipage}[b]{0.3\linewidth}
\begin{center}
\includegraphics[width = 2.5in]{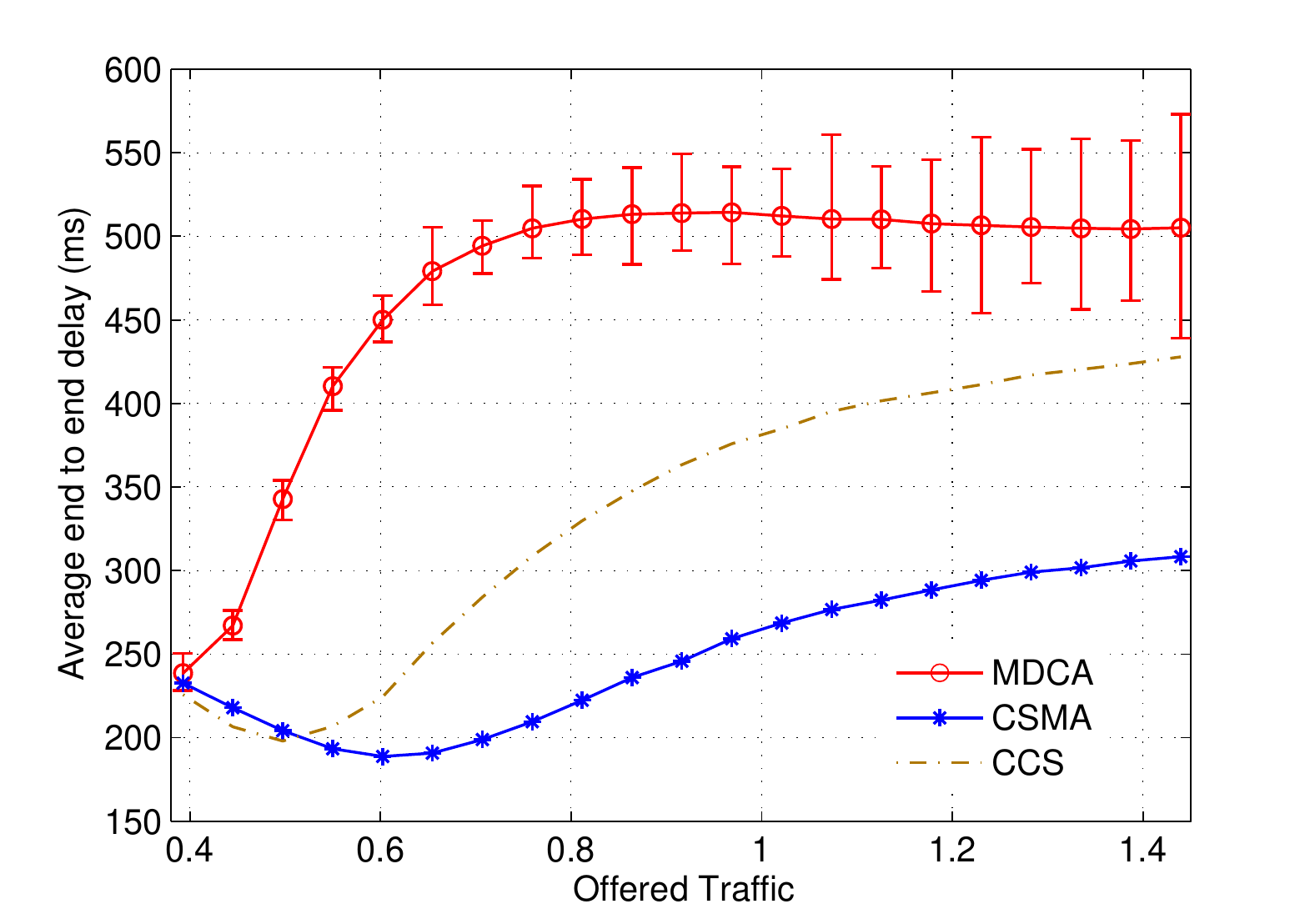}
\caption{Average end-to-end delay for different schemes (for $N = 20$, $M = 7$, $\eta = 4$).}
\label{delSO4dist}
\end{center}
\end{minipage}
\end{figure*}

\subsubsection{Performance of the MDCA scheme}

Figs.~\ref{pdrSO3dist}-\ref{delSO4dist} show the performance of the MDCA scheme. For comparison, we also consider the CSMA/CA protocol with no packet drops due to  the backoff limit or the retransmissions limit. This is indicated as CSMA2 in the figures. Fig.~\ref{pdrSO3dist} shows the packet delivery ratio (PDR) for different schemes. In the low congestion regime, the  MDCA scheme shows  performance similar to that of the CSMA2 scheme. When the MDCA scheme detects congestion, it starts using the TDMA slots during CFP according to the policy $\pi^*$. The use of CFP boosts the PDR of the nodes. As shown in Fig.~\ref{enSO3dist}, energy efficiency of channel access in terms of consumed energy per successfully transmitted packet per node in the network is also improved. The reason behind this is that transmitting packets during CFP avoids wasting energy in carrier sensing and retransmissions. As shown in Fig.~\ref{delSO3dist}, the price that the nodes have to pay for the improved PDR and energy efficiency is the increased end-to-end delay. One reason of this increased delay is, no packets are dropped because of limits of in number of backoffs or retransmissions. Another reason is, when a node transmits during an assigned TDMA slot, it has to wait during CAP. The CSMA/CA scheme shows the lowest end-to-end delay when dropping of packets is allowed during contention period. Tuning the MAC parameters would show better performance in low congestion region where CAP is long enough to transmit all packets using the CSMA/CA scheme~\cite{francesco}. However, we consider the superframe to be $384$ UBP long and a packet to be $6$ UBP long. The  performance of the CCS scheme is similar to  the performance of CSMA/CA scheme, because  in the CCS scheme the nodes are unable to tune the channel access parameters optimally  in a distributed fashion.

Figs.~\ref{pdrSO4dist}-\ref{delSO4dist} show the results for the scenario when $\eta$ is changed to 4. To achieve this, we double the CAP and the superframe period. With $K = 16$, the slot size is $T_{slot} = 48$ UBP. The proposed MDCA scheme has similar performance as in the case of $\eta = 2$. However, as shown in Fig.~\ref{pdrSO4dist}, the bandwidth utilization becomes worse as the slot size becomes larger. As the nodes with buffer level less than $\eta$ packets start using the TDMA slot, the packet delivery ratio does not improve because of bandwidth under-utilization. Therefore, a smaller slot size is desirable for the MDCA scheme.

\subsubsection{Performance of the MCCA scheme}
\begin{figure*}[t]
\begin{minipage}[b]{0.3\linewidth}
\begin{center}
\includegraphics[width = 2.5in]{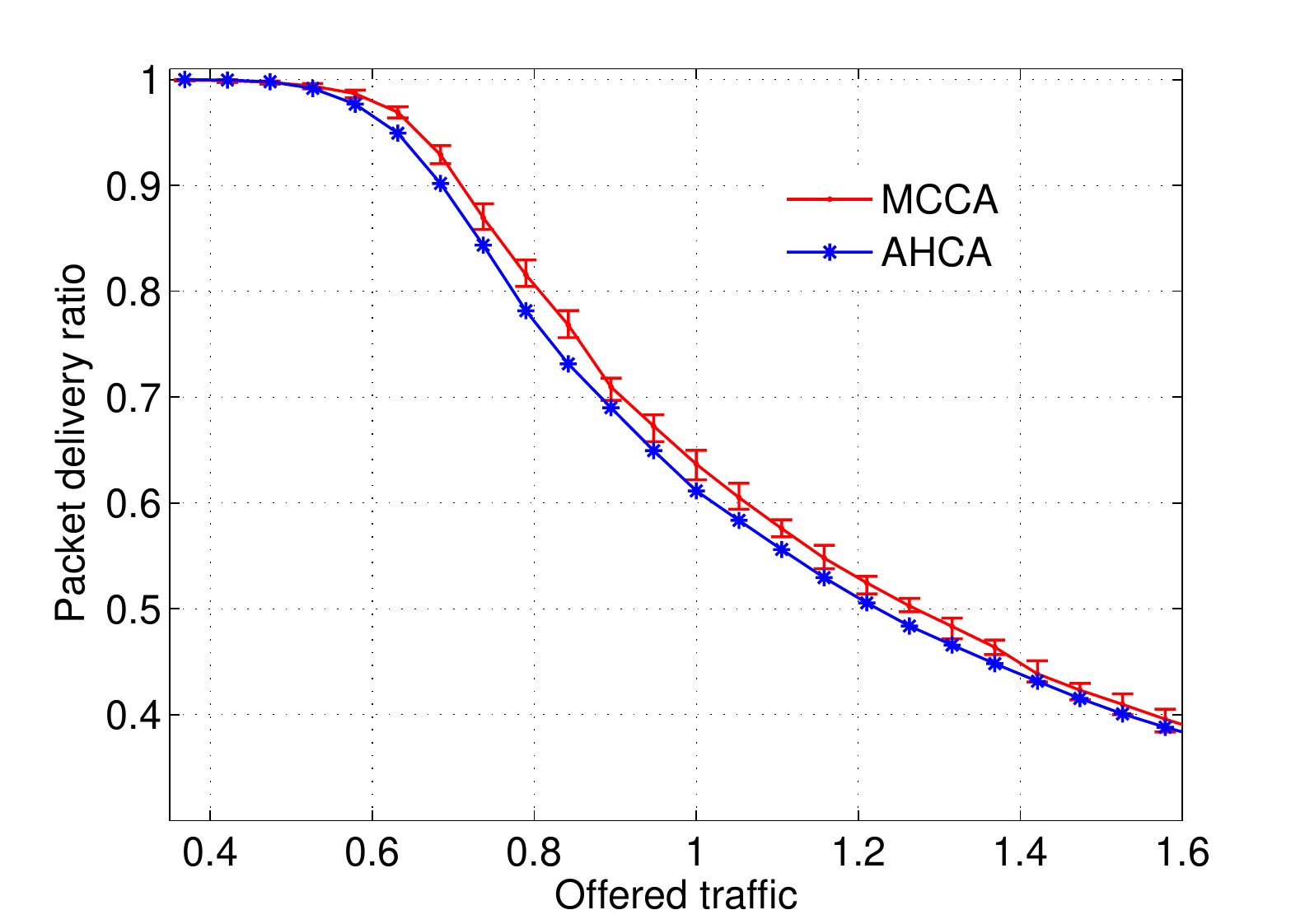}
\caption{Packet delivery ratio for different schemes (for $N = 20$, $M = 7$, $\eta = 2$).}
\label{pdrSO3cent}
\end{center}
\end{minipage}
\hspace{0.2cm}
\begin{minipage}[b]{0.3\linewidth}
\begin{center}
\includegraphics[width = 2.5in]{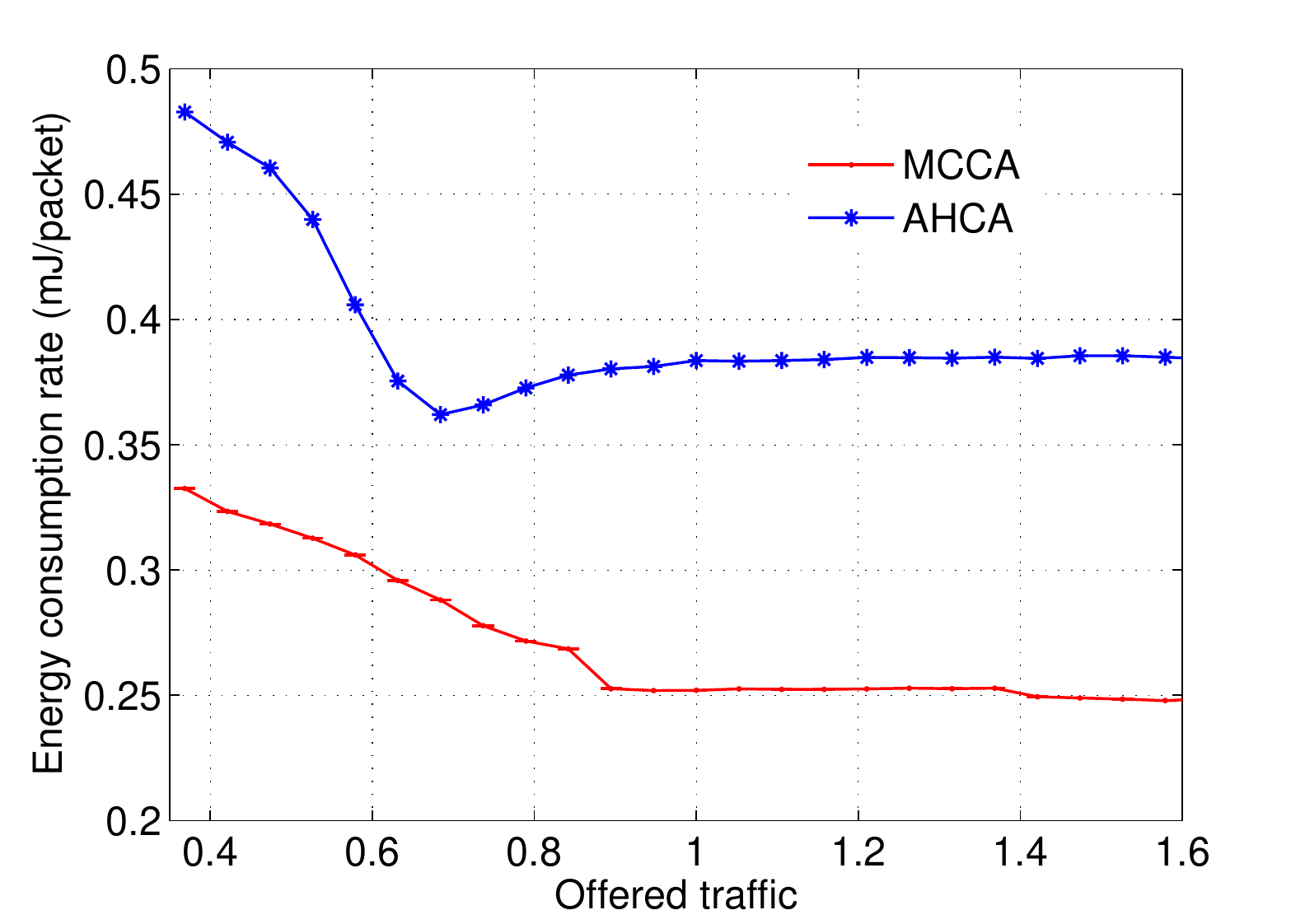}
\caption{Energy consumption rate for different schemes (for $N = 20$, $M = 7$, $\eta = 2$).}
\label{enSO3cent}
\end{center}
\end{minipage}
\hspace{0.2cm}
\begin{minipage}[b]{0.3\linewidth}
\begin{center}
\includegraphics[width = 2.5in]{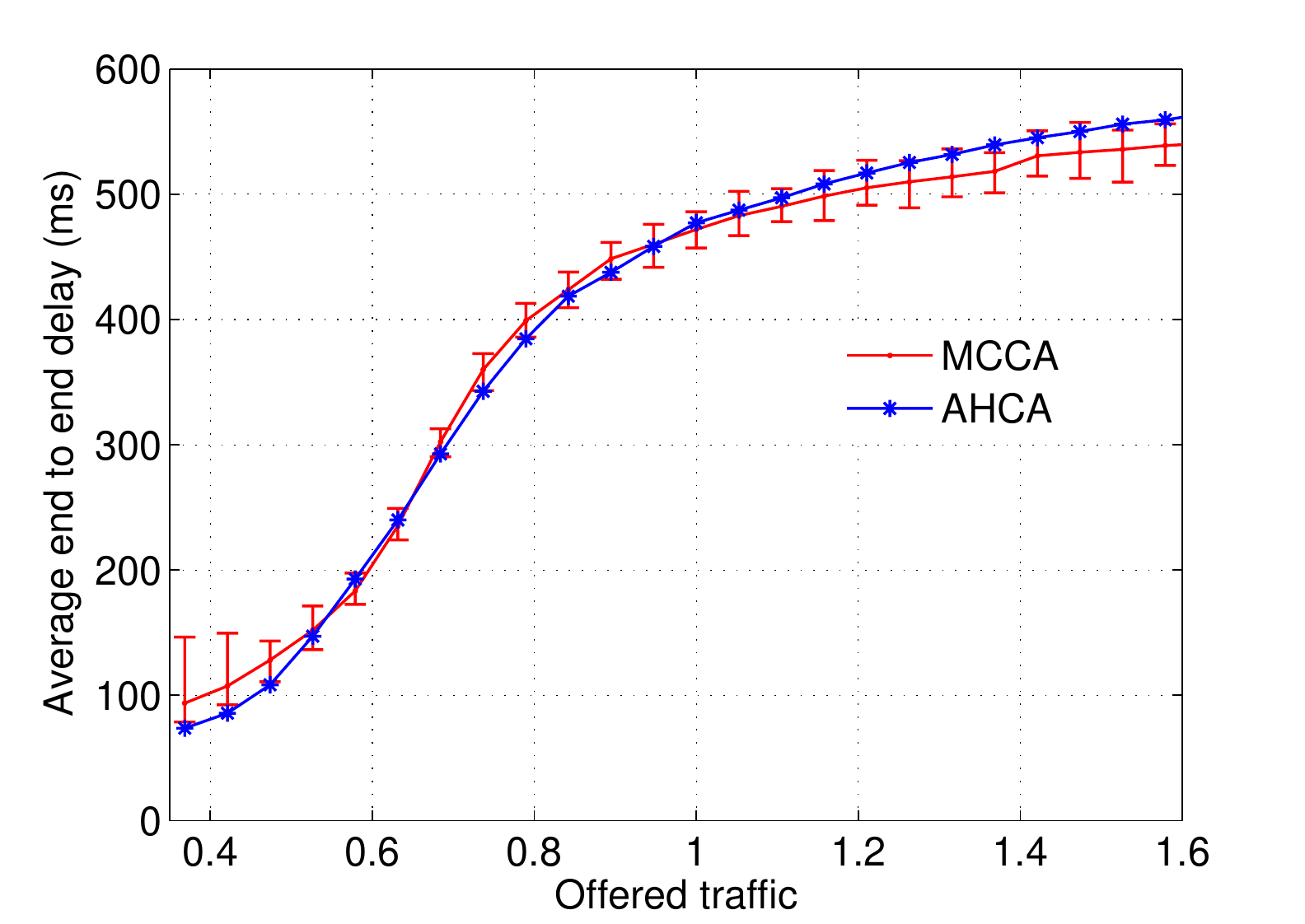}
\caption{Average end-to-end delay for different schemes (for $N = 20$, $M = 7$, $\eta = 2$).}
\label{delSO3cent}
\end{center}
\end{minipage}
\end{figure*}

\begin{figure*}[t]
\begin{minipage}[b]{0.3\linewidth}
\begin{center}
\includegraphics[width = 2.5in]{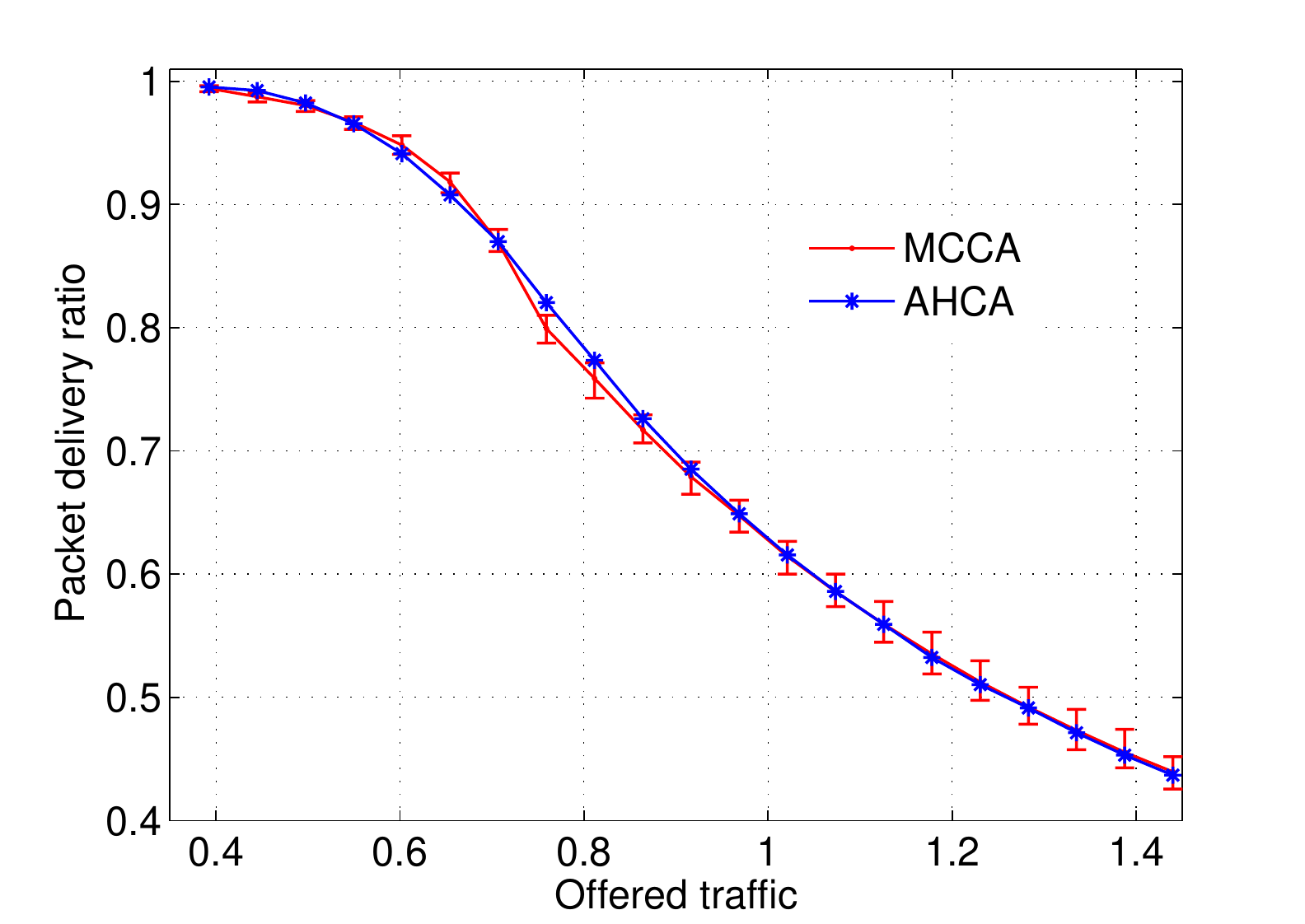}
\caption{Packet delivery ratio for different schemes (for $N = 20$, $M = 7$, $\eta = 4$).}
\label{pdrSO4cent}
\end{center}
\end{minipage}
\hspace{0.2cm}
\begin{minipage}[b]{0.3\linewidth}
\begin{center}
\includegraphics[width = 2.5in]{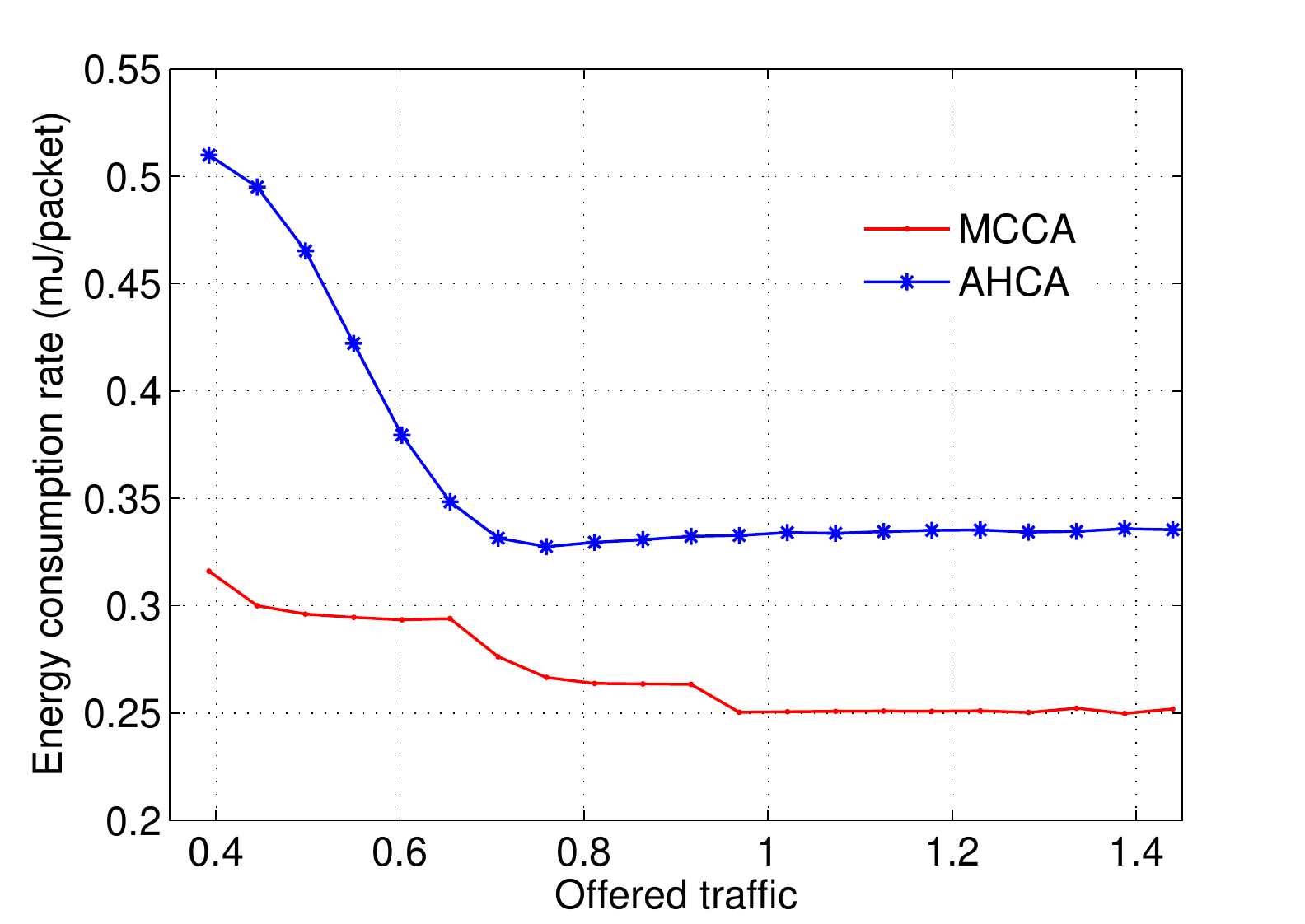}
\caption{Energy consumption rate for different schemes (for $N = 20$, $M = 7$, $\eta = 4$).}
\label{enSO4cent}
\end{center}
\end{minipage}
\hspace{0.2cm}
\begin{minipage}[b]{0.3\linewidth}
\begin{center}
\includegraphics[width = 2.5in]{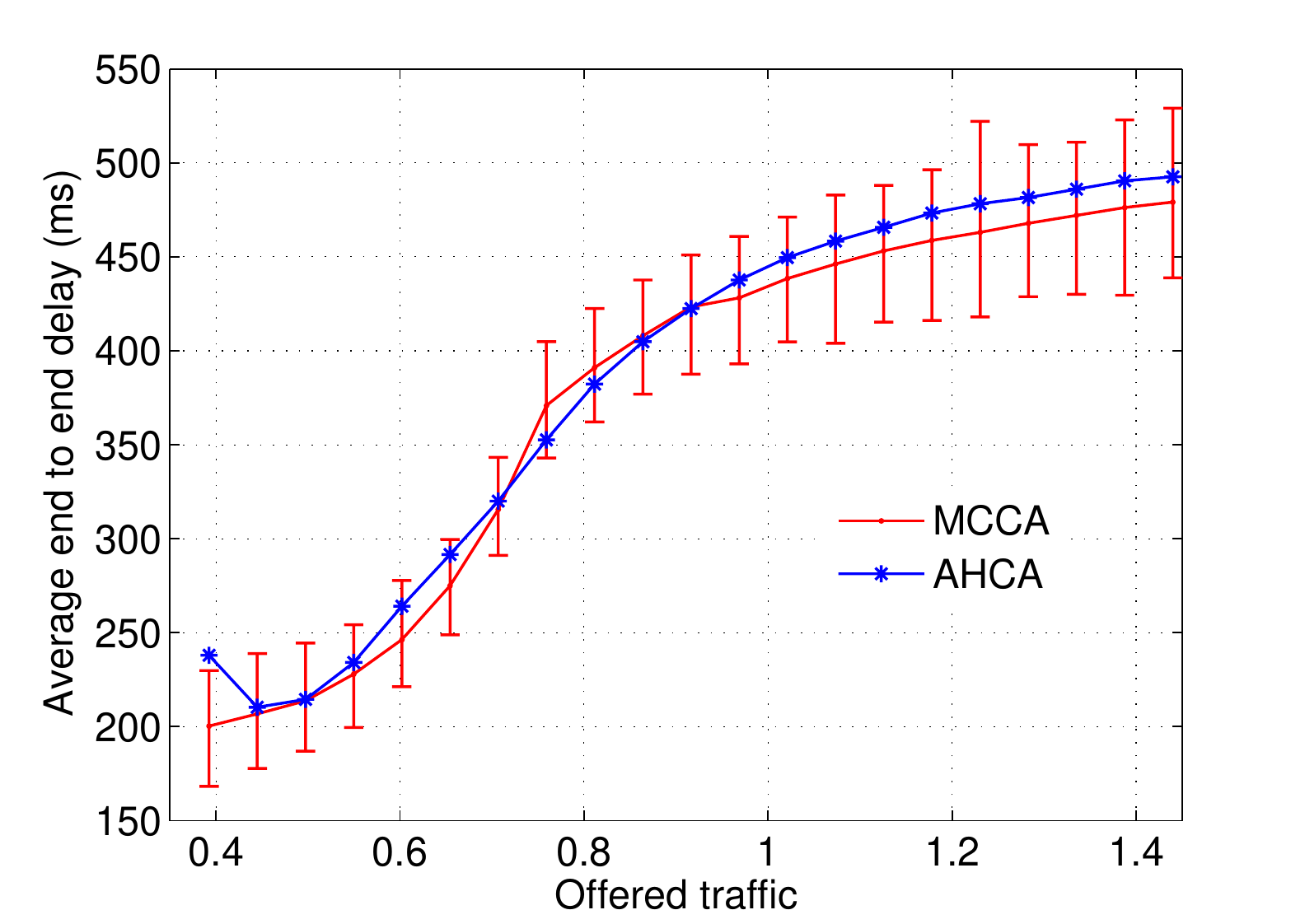}
\caption{Average end-to-end delay for different schemes (for $N = 20$, $M = 7$, $\eta = 4$).}
\label{delSO4cent}
\end{center}
\end{minipage}
\end{figure*}

Figs.~\ref{pdrSO3cent}-\ref{delSO4cent} show the performance results for the MCCA scheme. It is observed that both the MCCA and AHCA schemes have similar performances in terms of PDR and end-to-end delay. Figs.~\ref{enSO3cent} and~\ref{enSO4cent}  show that the MCCA scheme consumes less energy to transmit a packet successfully to the coordinator. This is due to the fact that, instead of letting all the nodes compete during CAP as in the AHCA scheme, the coordinator in the MCCA scheme schedules some nodes to go into the low-power mode (defer transmission) to maximize the total CAP throughput. However, this requires the coordinator to perform more computations to find out the list of the nodes that either transmit through CFP and CAP or defer transmissions.

By observing these figures we conclude that the MCCA scheme achieves a better performance than the other schemes. However, if the coordinator does not have the capability of processing the information of the traffic loads of all nodes, then the proposed MDCA scheme would be more desirable.

\begin{figure*}[t]
\begin{minipage}[b]{0.3\linewidth}
\begin{center}
\includegraphics[width = 2.5in]{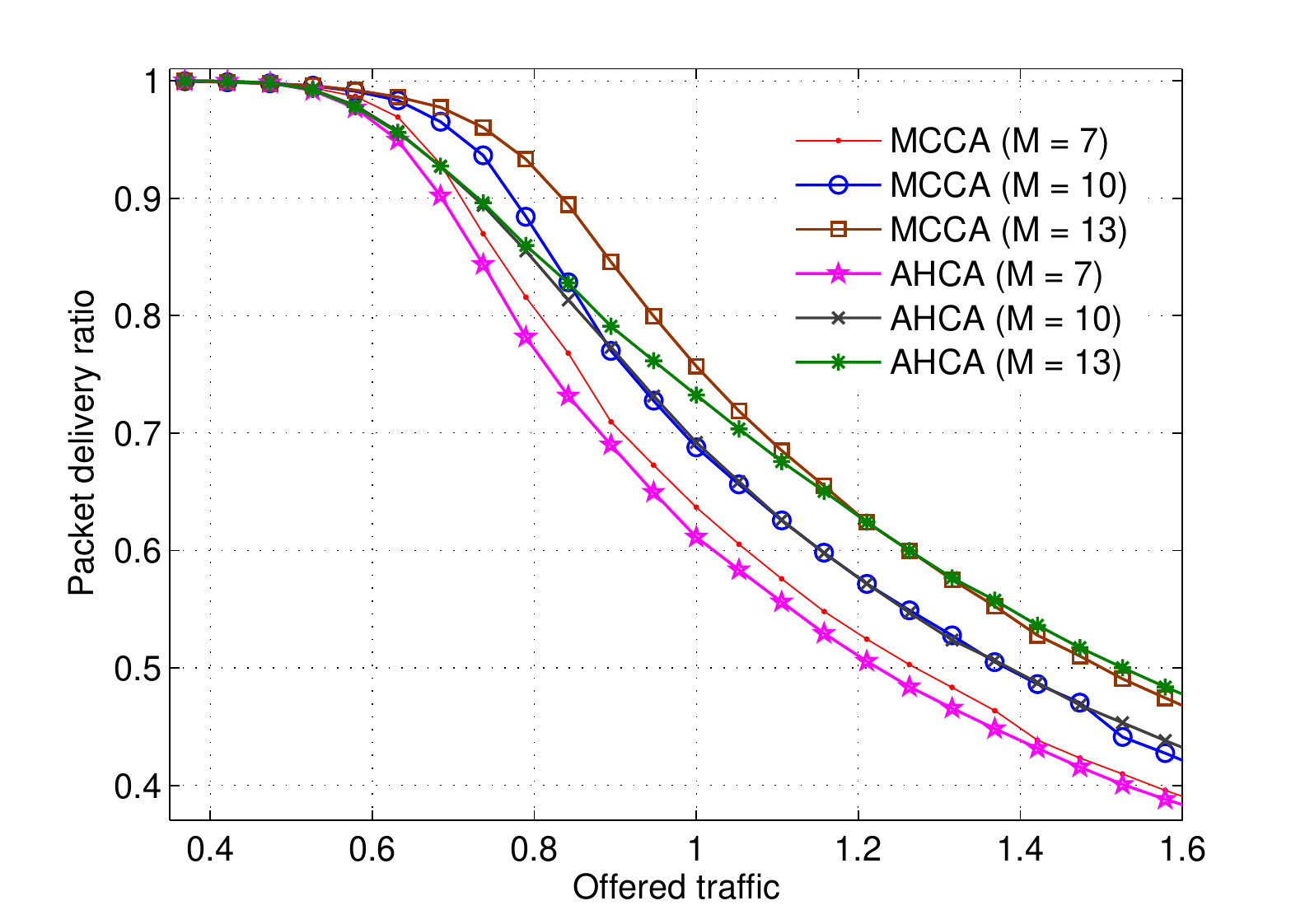}
\caption{Packet delivery ratio for different values of CFP length $M$ (for $N = 20$,  $\eta = 2$).}
\label{pdrslotvary}
\end{center}
\end{minipage}
\hspace{0.2cm}
\begin{minipage}[b]{0.3\linewidth}
\begin{center}
\includegraphics[width = 2.5in]{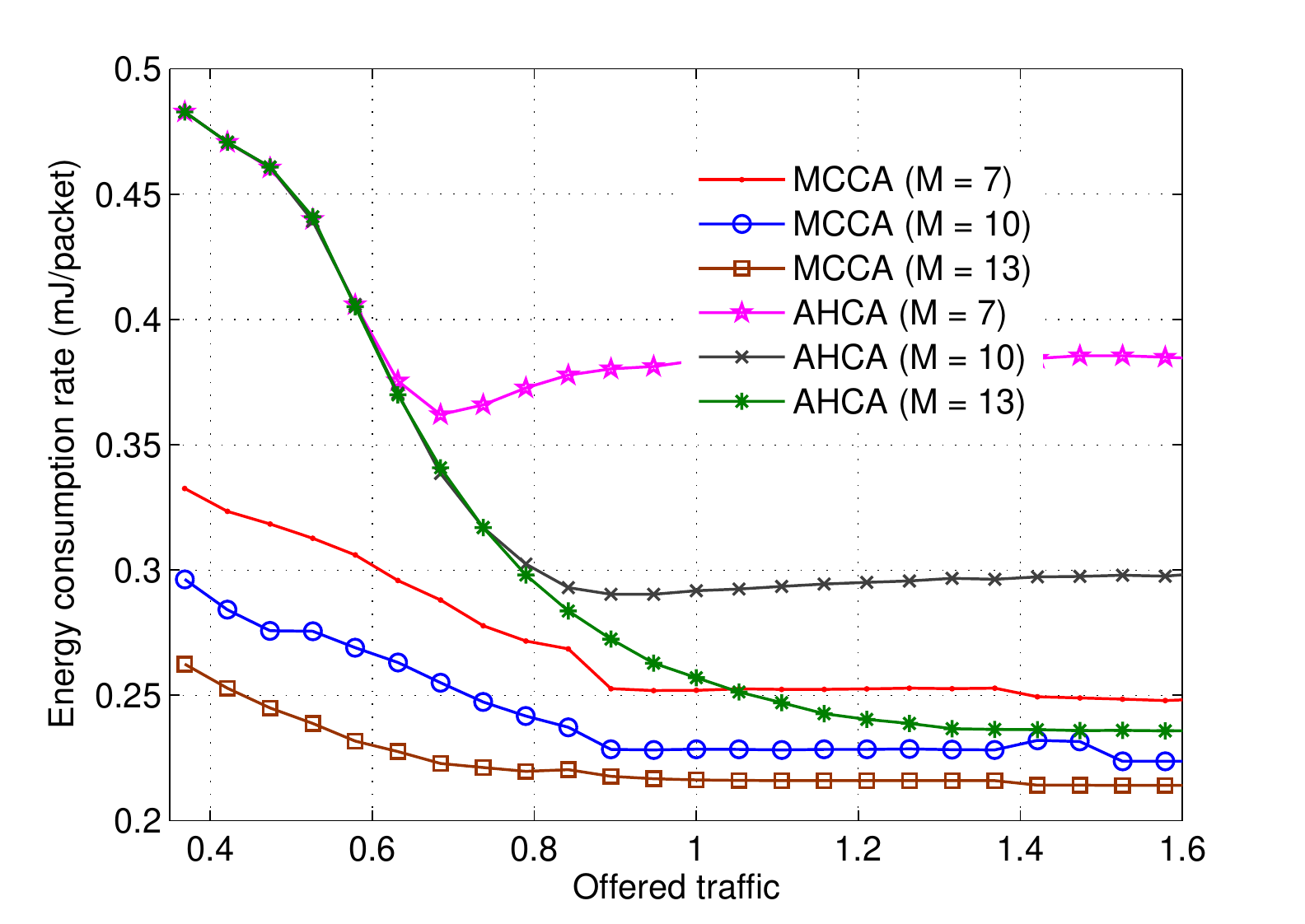}
\caption{Energy consumption rate for different values of CFP length $M$ (for $N = 20$,  $\eta = 2$).}
\label{enslotvary}
\end{center}
\end{minipage}
\hspace{0.2cm}
\begin{minipage}[b]{0.3\linewidth}
\begin{center}
\includegraphics[width = 2.5in]{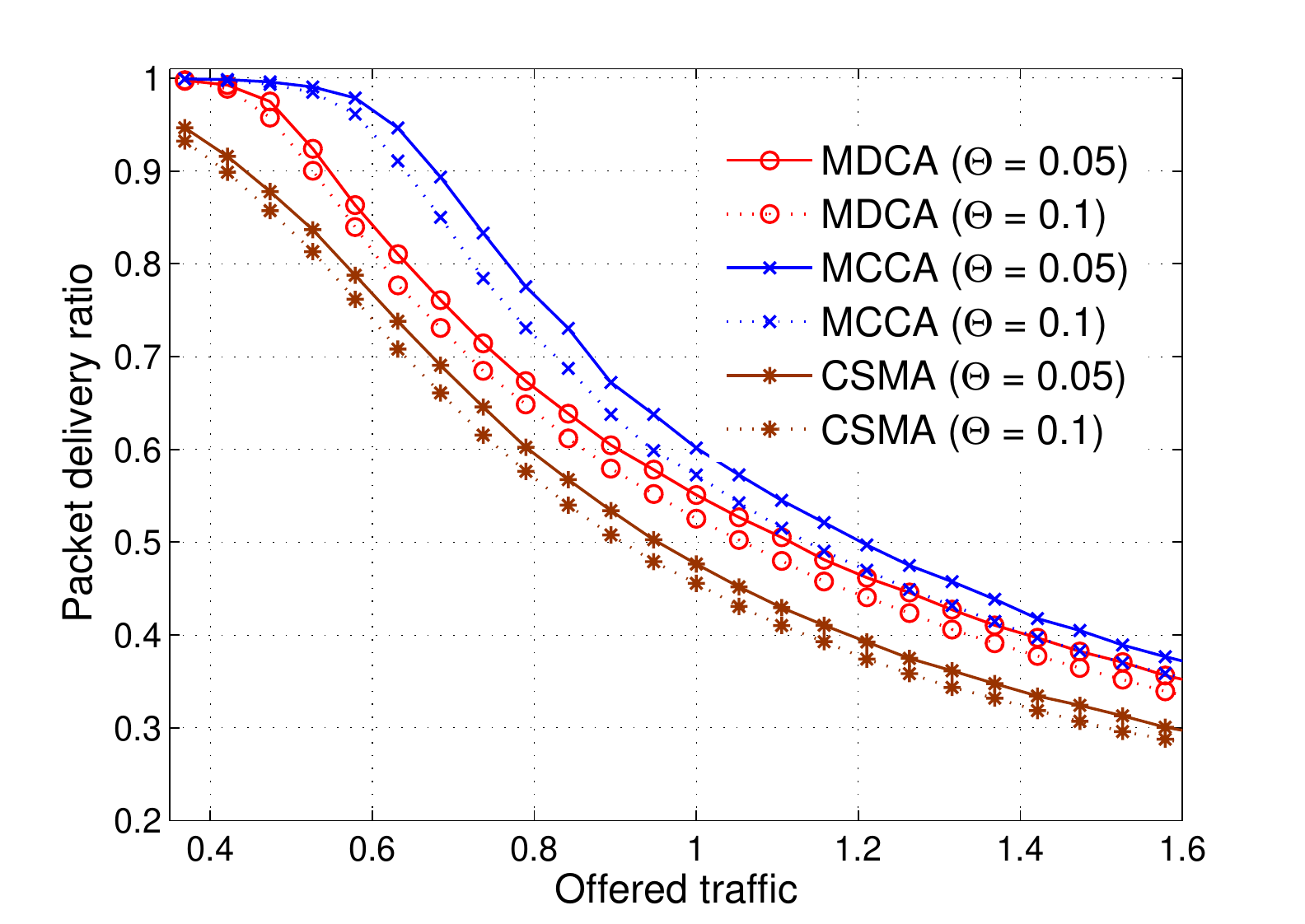}
\caption{Packet delivery ratio for different values of outage probability (for $N = 20$,  $M = 7$,  $\eta = 2$).}
\label{pdrout}
\end{center}
\end{minipage}
\end{figure*}

\subsubsection{Effect of number of time slots on the performance of the MCCA scheme}
We vary the number of TDMA slots ($M$) in the superframe. Note that the higher  the value of $M$, the smaller is the contention period. Also, $M = K$ means there is no contention period. For hybrid MAC, we need $M < K$. Figs.~\ref{pdrslotvary} and~\ref{enslotvary} show that, for both the MCCA and AHCA schemes, with increasing $M$ the nodes achieve a better performance. This is because of increased number of successful transmissions during CFP. At a lower traffic load, the proposed MCCA scheme achieves a better PDR than the AHCA scheme because of better bandwidth utilization. Also, as shown in Fig.~\ref{enslotvary}, sleep scheduling in the proposed MCCA scheme reduces the energy consumption.



\subsubsection{Effect of probability of outage on the performance of the MDCA and MCCA schemes}
We vary the probability that the packet is not received correctly at the coordinator (i.e., outage probability). In the simulation, $\Theta = 0.05$ means 5 out of 100 packets received by the coordinator from a node are erroneous. We assume that all the links between the nodes and the coordinator go into fading at the same time so that the hidden node collision does not have any adverse effect (i.e., $H = 0$). Fig.~\ref{pdrout} indicates that channel outage degrades the performance of the nodes because of increased congestion. The performance of the CSMA/CA scheme with $\Theta = 0.05$ is worse than the performance of the MDCA scheme with $\Theta = 0.1$. This shows that when the network becomes congested (because of increased traffic load and/or channel fading), the hybrid scheme performs better than the CSMA/CA scheme.

\begin{figure*}[t]
\begin{minipage}[b]{0.3\linewidth}
\begin{center}
\includegraphics[width = 2.5in]{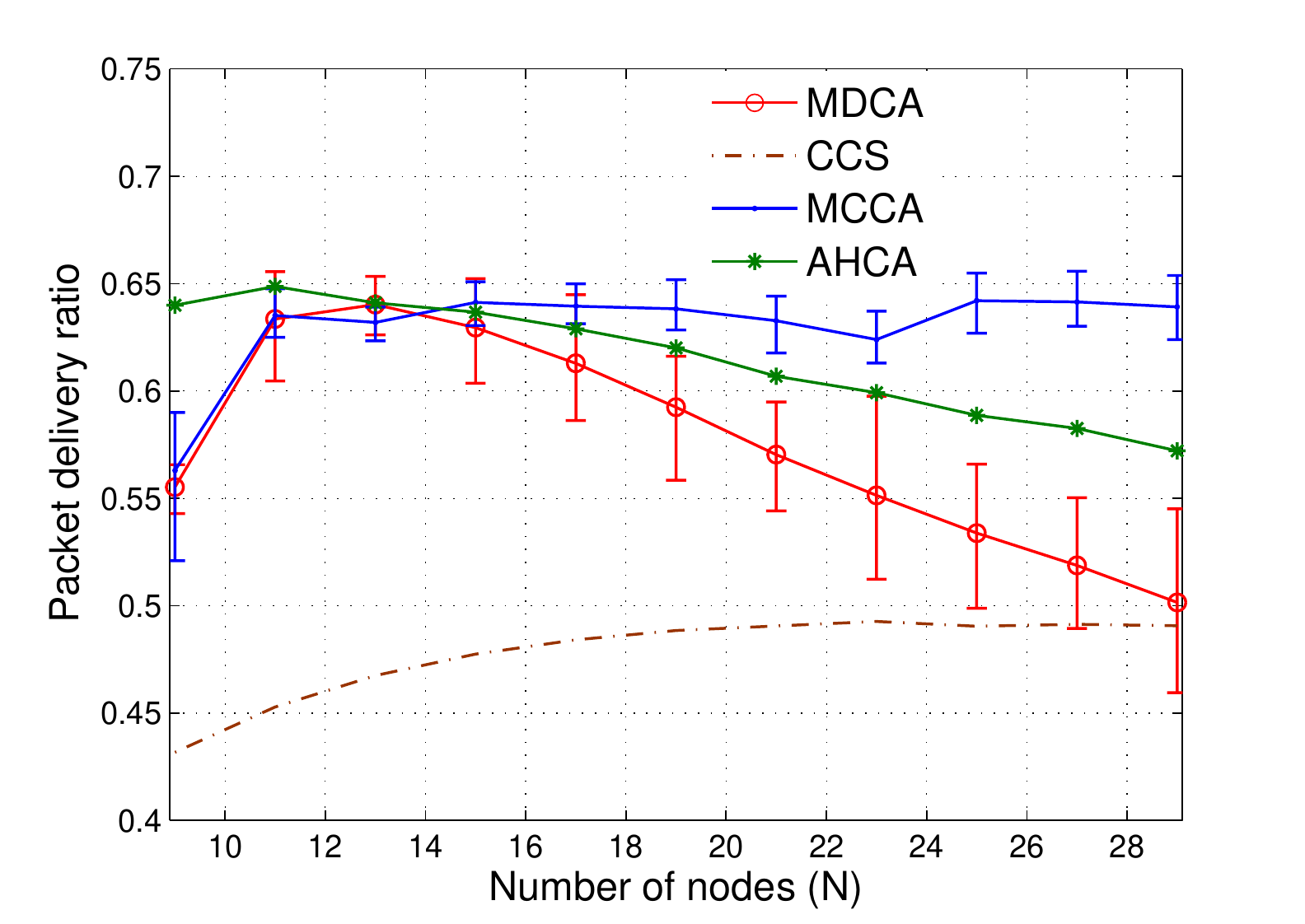}
\caption{Packet delivery ratio for different network size (for $M = 7$,  $\eta = 2$).}
\label{nvar_p}
\end{center}
\end{minipage}
\hspace{0.2cm}
\begin{minipage}[b]{0.3\linewidth}
\begin{center}
\includegraphics[width = 2.5in]{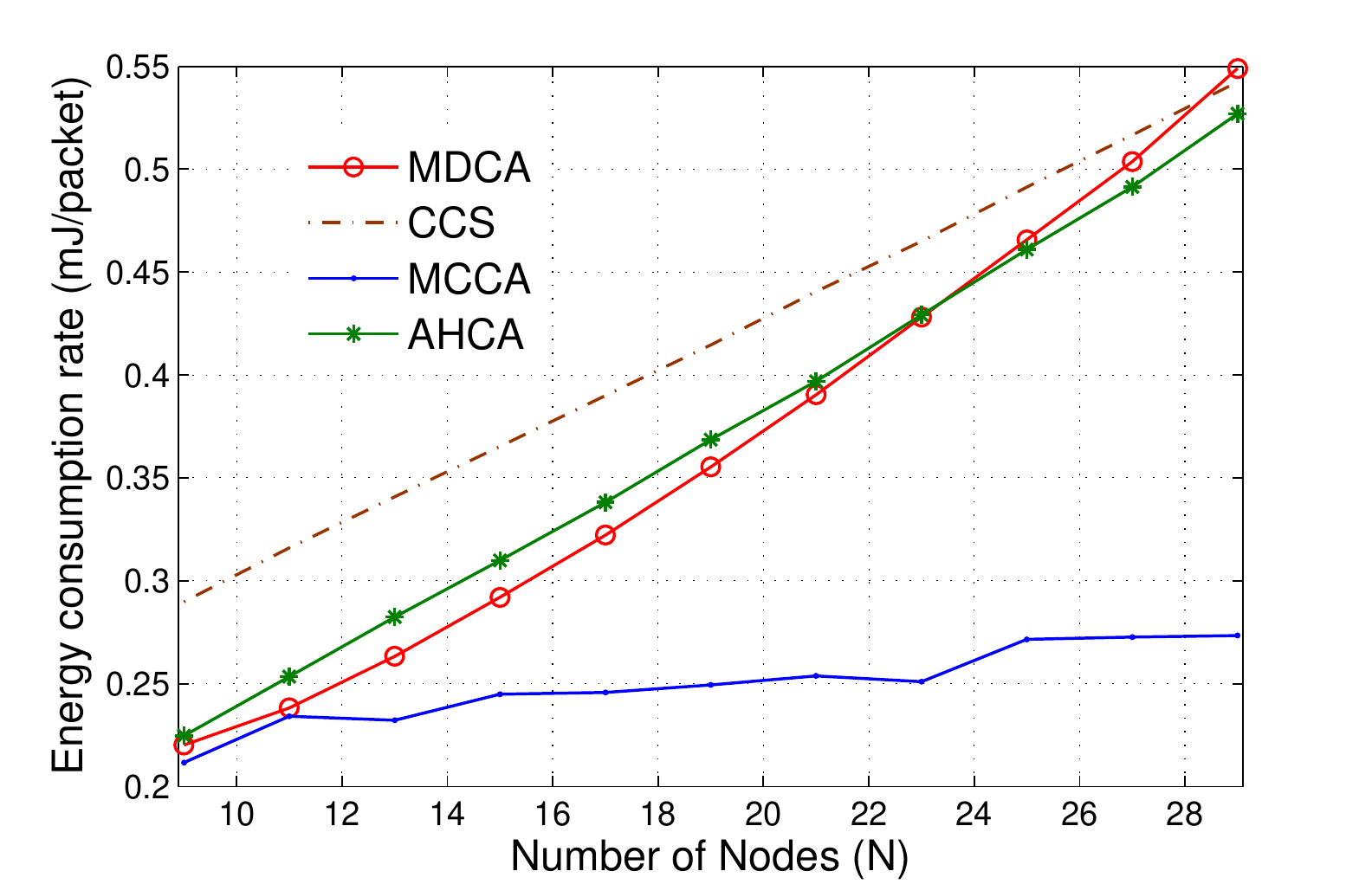}
\caption{Energy consumption rate for different network size (for $M = 7$,  $\eta = 2$).}
\label{nvar_en}
\end{center}
\end{minipage}
\hspace{0.2cm}
\begin{minipage}[b]{0.3\linewidth}
\begin{center}
\includegraphics[width = 2.5in]{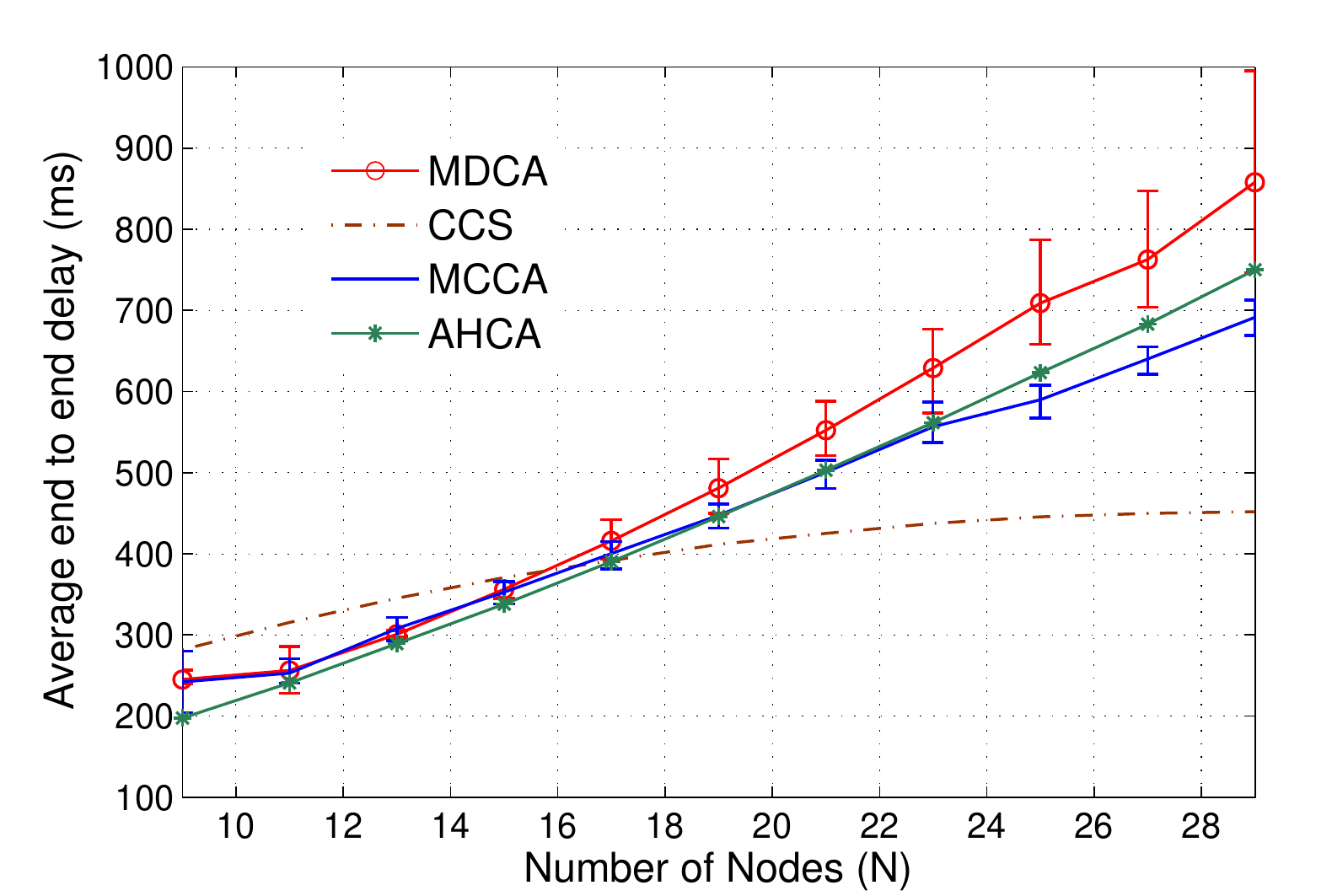}
\caption{Average end-to-end delay for different network size (for $M = 7$,  $\eta = 2$).}
\label{nvar_del}
\end{center}
\end{minipage}
\end{figure*}

\subsubsection{Effect of network size on the performances of the MDCA and MCCA schemes}

We vary the number of nodes ($N$) in the network. The packet arrival rate of a node is considered to be $\lambda = \frac{T_{sf}+T_{beacon}}{T_{tx} N}$. The packet rate is enough to push the network into congestion region. Fig.~\ref{nvar_p} shows a comparison among different schemes in terms of the packet delivery ratio. Since the total traffic in the network is  inversely proportional to the network size $N$, the packet delivery ratio per node is almost flat for the CSMA/CA, CCS, and centralized schemes. However, the performance of the nodes in the MDCA scheme is dependent on the bandwidth utilization. The nodes require a sufficiently long  contention period to transmit the TDMA slot reservation request successfully. For a higher number of nodes $N$, the contention period becomes more congested. Eventually, the number of successful requests for TDMA slots decreases and the bandwidth utilization becomes worse. This is the reason why the performance of the MDCA scheme degrades as the network size ($N$) increases. Therefore, for an efficient operation of the MDCA scheme,  the contention period and the number of TDMAs slot need to selected appropriately.

As shown in Fig.~\ref{nvar_en}, the energy consumption rate grows almost linearly in all the schemes except the MCCA scheme. The reason for linear increase is that the throughput of a node saturates for higher network size but the energy consumption increases due to higher number of retransmissions and carrier sensing. However, in the MCCA scheme, scheduling of the nodes to go into low power mode makes the ratio of energy consumption to the throughput remain at almost the same level.  The tradeoff between the average end-to-end delay and energy consumption is  shown in Fig.~\ref{nvar_del}.

\subsubsection{Performances of the MDCA and MCCA schemes under heterogeneous traffic}

\begin{figure*}[t]
\begin{minipage}[b]{0.3\linewidth}
\begin{center}
\includegraphics[width = 2.3in]{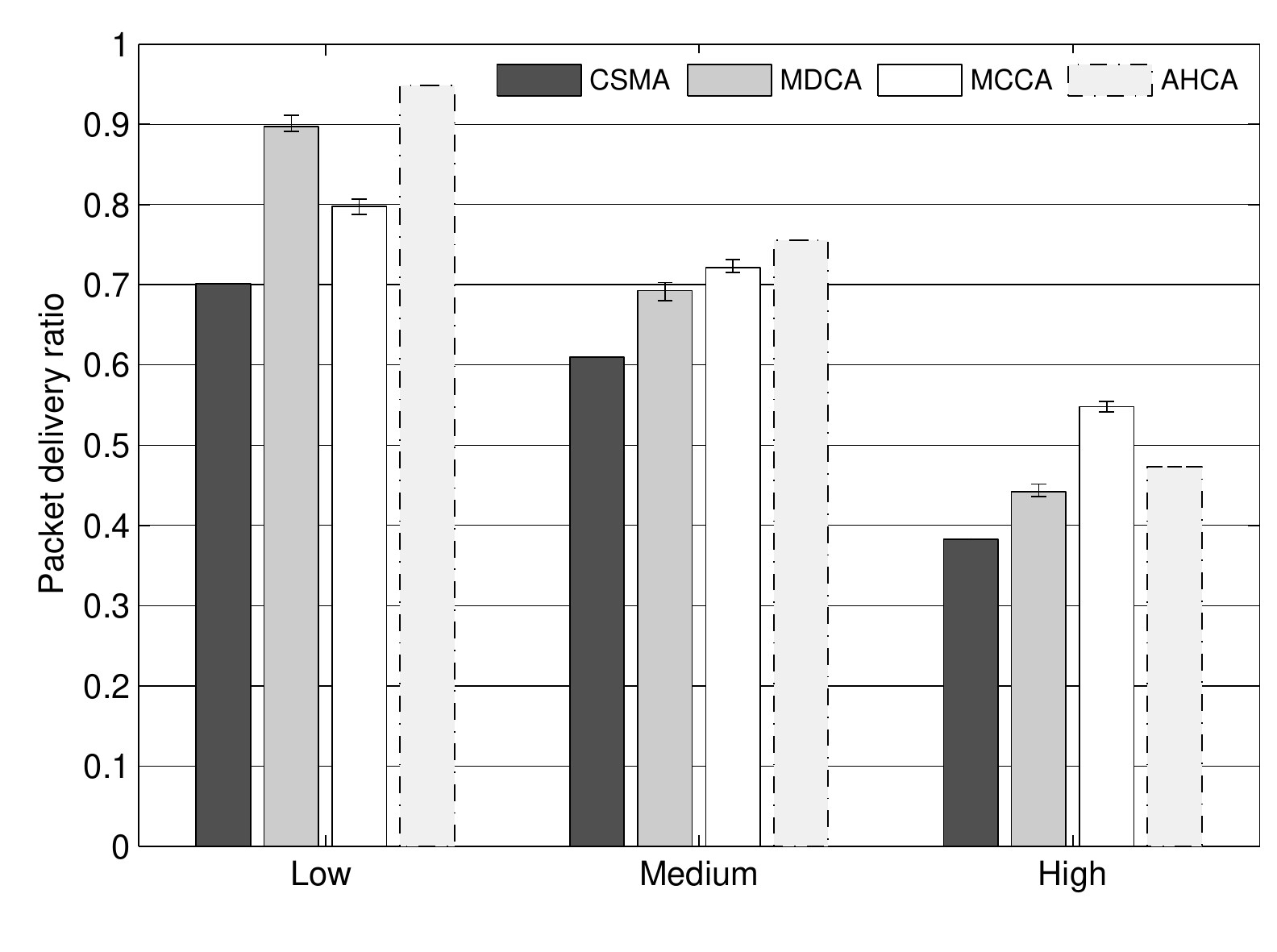}
\caption{Packet delivery ratio for different groups (for $N' = 6$ nodes, $M = 7$,  $\eta = 2$).}
\label{pdrN18het}
\end{center}
\end{minipage}
\hspace{0.2cm}
\begin{minipage}[b]{0.3\linewidth}
\begin{center}
\includegraphics[width = 2.3in]{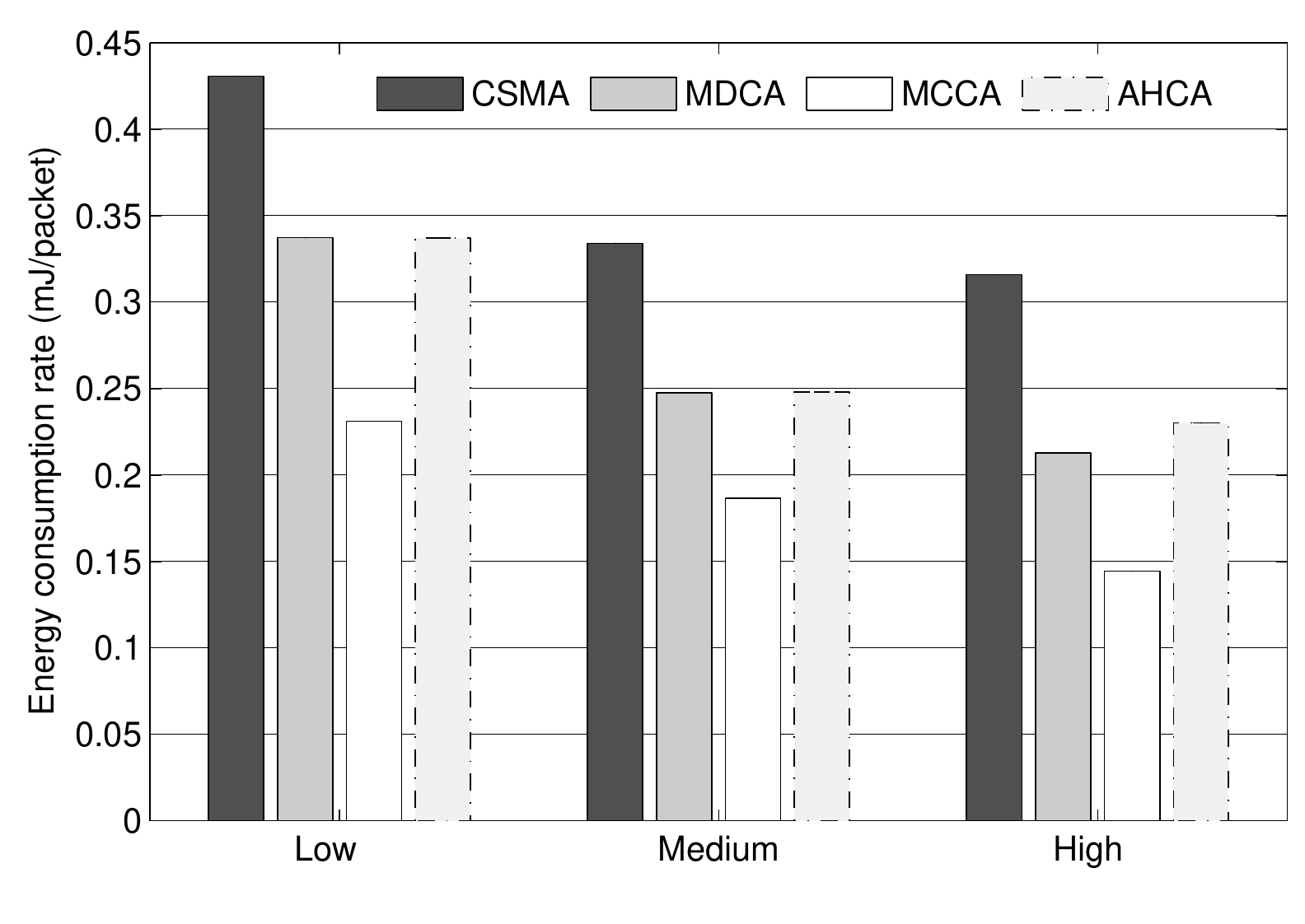}
\caption{Average energy consumption rate for different groups (for $N' = 6$ nodes, $M = 7$,  $\eta = 2$).}
\label{enn18het}
\end{center}
\end{minipage}
\hspace{0.2cm}
\begin{minipage}[b]{0.3\linewidth}
\begin{center}
\includegraphics[width = 2.3in]{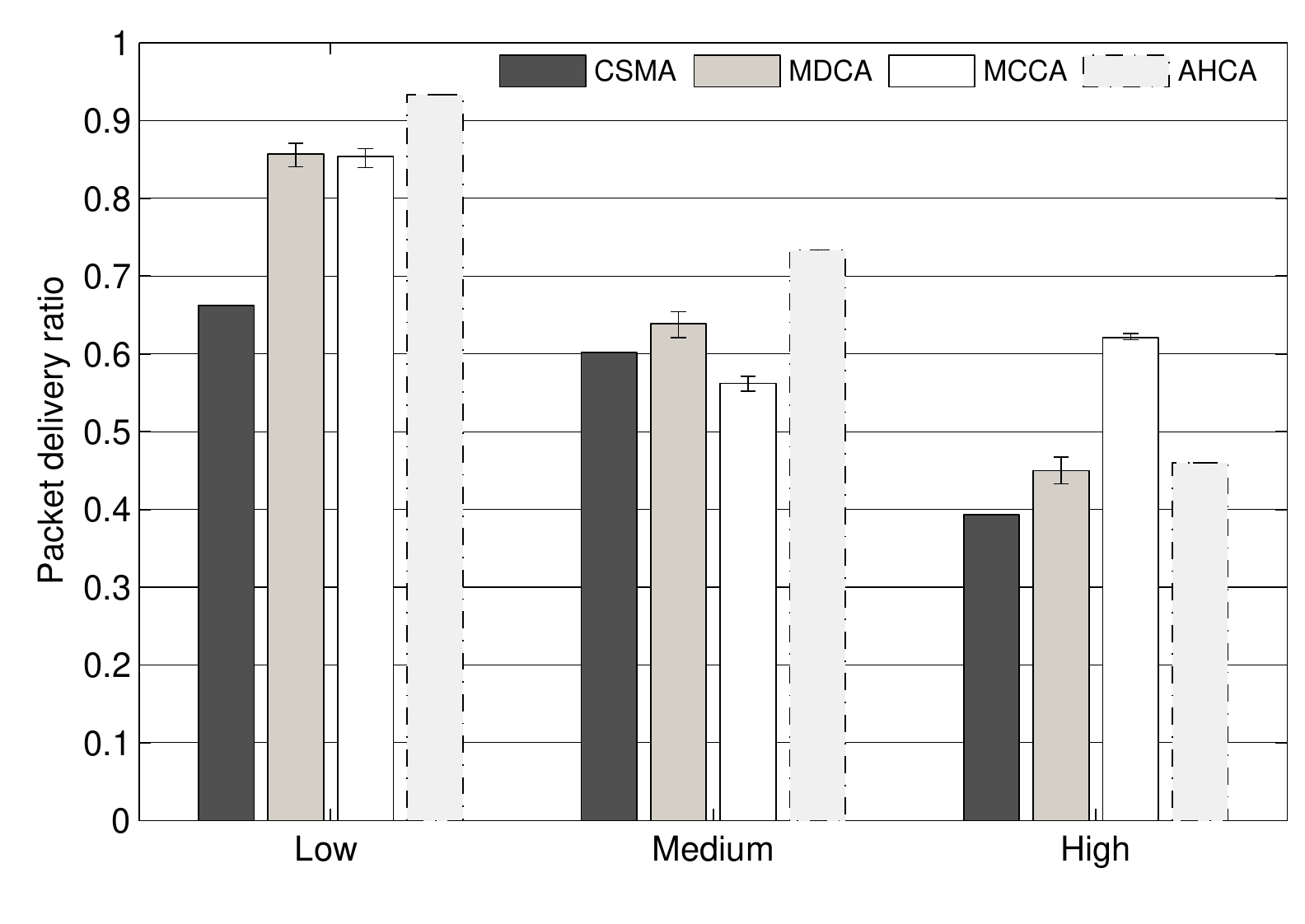}
\caption{Packet delivery ratio for different groups (for $N' = 7$ nodes, $M = 7$,  $\eta = 2$).}
\label{pdrN21het}
\end{center}
\end{minipage}
\end{figure*}

We divide the $N$ nodes into three groups based on their traffic, namely, the low rate group, the medium rate group and the high rate group. The size of each group is $N' = N/3$. We investigate the performance of the nodes with heterogeneous traffic in the network. The nodes in the three  groups have packet arrival rates of $\lambda_{low} = 0.15\frac{T_{sf}+T_{beacon}}{N'T_{tx}}$, $\lambda_{medium} = 0.30\frac{T_{sf}+T_{beacon}}{N'T_{tx}}$, and  $\lambda_{high} = 0.55\frac{T_{sf}+T_{beacon}}{N'T_{tx}}$, respectively. Figs.~\ref{pdrN18het}-\ref{pdrN21het} show that the MDCA scheme performs better than the CSMA/CA scheme in the heterogeneous traffic scenario as well. As the lower rate nodes mostly transmit during CAP, they have higher energy consumption rate. Even though the offered traffic rates of the nodes are heterogeneous, the transmission policy $\pi*$ developed for the saturation region  in the proposed schemes works well. Similarly, the MCCA scheme is better for higher rate nodes because they mostly use the slots during CFP.

\section{Related Work}


One of the pioneering works that deals with switching between contention access and contention free (i.e., TDMA) access was presented in~\cite{hybrid}. The model, which was designed for optical networks, switches contention access to contention free access when collision rate is high. Another related work can be found  in~\cite{Ephrem} where each node in the network randomly selects a favored slot from the next window of  slots. Some other work on hybrid MAC include those in~\cite{cheng},~\cite{bin} and~\cite{zmac}. In~\cite{cheng}, the access point polls a node and the polled node transmits without contention while the rest of the nodes start the contention process. This method is not energy-efficient since the nodes have to overhear every packet. The model presented in~\cite{bin} offers contention access, scheduled time-division multiple access (TDMA), and polling-based TDMA. Based on channel status and traffic request, the coordinator maintains the size of contention period and slot allocations.
The authors in~\cite{zmac} proposed the concept of hybrid CSMA/CA and TDMA schemes in static TDMA-based wireless networks. The nodes in the network are allocated conflict-free TDMA slots.  If the slot owner has no packet to transmit, then non-slot owners compete to get access to the slot using CSMA/CA.  The work in~\cite{Wei} presented an improvement on the local framing scheduling model of~\cite{zmac}. The work in~\cite{Rana} considered bandwidth-aware TDMA slot allocation. Because of their static nature, these models have scalability problem.

The performance of  the hybrid MAC protocol in the IEEE 802.15.4 standard~\cite{stand} has been analyzed in the literature~\cite{bharattitb}, \cite{sheu}. The work in~\cite{bharattitb} presented a general discrete-time Markov chain model  taking into account the CSMA/CA and GTS-based transmissions together in a heterogeneous traffic scenario and non-saturated condition. The authors in~\cite{sheu} used GTS to cope with the hidden node collision problem in the IEEE 802.15.4-based personal-area networks.
The authors in~\cite{Zhang2} analyzed the performance of contention/reservation interleaved hybrid MAC using soft reservation where owner can release the unused reserved time.

Some work in the literature dealt with sleeping mechanisms in the IEEE 802.15.4 MAC~\cite{jurdak},~\cite{khanafer},~\cite{xiao}. In~\cite{jurdak}, the authors implemented  the IEEE 802.15.4-based RFID nodes considering only the non-beacon-enabled mode.
To save energy, such a node stays in sleep mode until it has data to transmit or the RFID tag is triggered to receive data.
In~\cite{khanafer},  to save energy, the authors proposed a strategy to force the nodes to go to sleep mode after each successful transmission. This strategy also helps reduce collision during CAP.  In~\cite{xiao}, the authors presented a Markov-based model taking into account the sleep mechanism of the IEEE 802.15.4. All of these work  focused only on contention-based channel access.

The use of contention access and TDMA access can be also optimized to enhance the performance (in terms of throughput and/or delay and/or energy) of the network. A knapsack model was presented in~\cite{gts} taking into account the bandwidth demand from nodes to allocate the guaranteed time slots to improve the throughput performance in the IEEE 802.15.4 networks. The authors in~\cite{bharatglb} presented a Markov decision process model to make the best use of CSMA/CA and guaranteed time slots to enhance the throughput and energy performance of the IEEE 802.15.4 networks. However, the problem of under-utilization of TDMA slots degrades the network performance.

Some work~(e.g., \cite{Gilani},~\cite{Zhuo}) considered the queue length-based TDMA slot allocation scheme to enhance the throughput and energy performance of hybrid random access and TDMA-based networks. In the model presented in~\cite{Gilani}, the coordinator allocates slots to the nodes according to their queue lengths to improve the throughput and energy efficiency performances in the IEEE 802.15.4 networks. Similarly, in~\cite{Zhuo}, the coordinator takes the queue lengths of nodes as the indicator of traffic.
The allocation of slots in these work is similar to that of LQF scheduling method which is considered to be throughput maximal~\cite{Lecon}.

In the literature, MDP-based models have been used for optimizing channel access in a wireless network  \cite{Sikdar}, \cite{Liu}, \cite{Angel}, and \cite{Phan}. In~\cite{Sikdar}, the authors developed an MDP model for wireless body-area sensor networks to balance the tradeoff between energy consumption and packet error rate.
In~\cite{Liu}, the authors presented a reinforcement learning-based solution for the MDP model to maximize the throughput and energy-efficiency in a wireless sensor network.
The authors in~\cite{Angel} considered the slotted ALOHA random access protocol and proposed an MDP model to take the optimal action. Based on the state (i.e., idle or backlogged), users choose their optimal transmit power and retransmission probability at the beginning of each time slot. The model was also extended for  the general case where users do not have the information about the backlogged users.
In~\cite{Phan}, the authors developed an MDP model for the transmission strategy of users in the IEEE 802.11 MAC-based wireless sensor networks. Using MDP, the users decide whether to transmit or defer transmission depending on the state (i.e., channel state, idle or active state of node) to minimize energy-consumption and frame error rate. In~\cite{van}, the authors  presented a post-decision state to cope with unknown traffic and channel condition in the network.

To the best of our knowledge, the problem of efficient channel access in a hybrid CSMA/CA-TDMA framework considering energy consumption, packet delivery ratio, the hidden node collision problem as well as traffic heterogeneity has not been addressed in the literature.
The  MDP-based transmission strategies presented in this paper consider the above aspects and handle  congestion in the network in a  way to improve the channel access performance in terms of packet delivery ratio and energy consumption.

\section{Conclusion}

We have proposed two MDP-based channel access schemes, namely, the MDCA and MCCA schemes,  to improve the performance of hybrid CSMA/CA and TDMA-based single hop wireless networks (e.g., IEEE 802.15.4-based networks). These schemes are useful to cope with congestion in the network which may result due to increased traffic load and/or channel fading. We have extended the performance analysis models for these schemes to consider channel fading and hidden node collisions. The performance evaluation results have shown that the proposed MDCA scheme improves network performance by detecting congestion in an intelligent way. The results show that the MCCA scheme is superior  but it requires information of packet arrival rate and instantaneous buffer level at all the network nodes. The proposed MCCA scheme is better than the existing hybrid CSMA/TDMA scheme in terms of energy consumption but it requires more computational effort. The proposed MDCA scheme is better (compared to the traditional schemes) when the information of traffic of all the nodes is unknown to the coordinator. Also, the MDCA scheme requires a shorter beacon frame because it does not contain information on the actions and the assignment of TDMA slots to the nodes. The MDCA scheme can be enhanced by using a de-centralized partially observable Markov decision process (DecPOMDP)  modeling approach to consider  non-Poisson traffic scenarios. 



\begin{thebibliography}{1}
\vspace{-0.25mm}
\bibitem{stand}
IEEE Standard for Information Technology Part 15.4: Wireless Medium Access Control (MAC) and Physical Layer (PHY) Specifications for Low-Rate Wireless Personal Area Networks (LR-WPANs),{ \em IEEE Standard 802.15.4 Working Group Std.}, 2006.

\bibitem{hybrid}
H.~I.~Liu and J.~D.~Wu, ``A Hybrid MAC Protocol for HFC Networks,'' in {\em Proc. of IEEE ICC'98}. vol.~ 2, pp.~859--863, Atlanta, 1998.

\bibitem{puterman}
 M.~L.~Puterman, {\em Markov Decision Processes: Discrete Stochastic Dynamic Programming}. New Jersey: John Wiley and Sons, Inc., 1998.

\bibitem{bharatlet}B.~Shrestha, K.~W.~Choi, and E.~Hossain, ``A Dynamic Time Slot Allocation Scheme for Hybrid CSMA/TDMA MAC Protocol,'' { \em IEEE Wireless Communications Letters}, to appear.

\bibitem{bharatglb} B.~Shrestha, E.~Hossain, K.~W.~Choi, and S.~Camorlinga, ``A Markov Decision Process (MDP)-Based Congestion-Aware Medium Access Strategy for IEEE 802.15.4,'' in {\em Proc. of IEEE Global Communications Conference (Globecom 2011)}, pp.~1--5, Houston, TX, USA, 5-9 December 2011.

\bibitem{Adelson} R.~M.~Adelson,``Compound Poisson distribution,'' {\em Operation Research}, vol.~17, no.~1, pp.~73--75, Mar.~1966.

\bibitem{patro} R.~K.~Patro, M.~Raina, V.~Ganapathy, M.~Shamaiah, and C.~Thejaswi, ``Analysis and Improvement of Contention Access Protocol in IEEE 802.15.4 Star Network,'' in {\em Proc. of IEEE  Conf. on Mobile Adhoc and Sensor Systems (MASS)}, pp.~1--8, Pisa, October 2007.

\bibitem{park1} P.~Park, P.~D.~Marco, P.~Soldati, C.~Fischione, and K.~H.~Johansson, ``A Generalized Markov Chain Model for Effective Analysis of Slotted IEEE 802.15.4,'' in {\em Proc. IEEE Int. Conf. on Mobile Ad-hoc and Sensor Systems}, pp.~130--139, Macau, 2009.

\bibitem{bharattitb} B.~Shrestha, E.~Hossain, and S.~Camorlinga, ``IEEE 802.15.4 MAC with GTS Transmission for Heterogeneous Devices with Application to Wheelchair Body-Area Sensor Networks,'' { \em IEEE Trans. on Information Technology in Biomedicine}, 15(5), pp.~767--777, September 2011.

\bibitem{park} P.~Di~Marco, P.~Park, C.~Fischione, and K.~H.~Johansson, ``Analytical Modeling of Multi-hop IEEE 802.15.4 Networks,'' { \em IEEE Transactions on Vehicular Technology}, 61(7), pp.~3191--3208, September 2012.


\bibitem{littman}
L.~P.~Kaelbling,  M.~L.~Littman, and A.~W.~Moore, ``Reinforcement Learning: A Survey,'' {\em Journal of Artificial Intelligence Research}, vol. 4, pp.~237--285, 1996.


\bibitem{Lecon} M.~Leconte,N.~Jian, R.~Srikant, ``Improved Bounds on the Throughput Efficiency of Greedy Maximal Scheduling in Wireless Networks,'' {\em IEEE/ACM Transactions on Networking}, 19(3), pp.~709--720, June 2011.

\bibitem{yazdan}
K.~Y.~Yazdandoost and K.~Sayrafian-Pour, ``Channel Model for Body Area Network (BAN),'' { \em IEEE 802.15 Working Group Document, IEEE P802.15-08-0780-09-0006}, April 2009.


\bibitem{chipcon} Chipcon CC2420 2.4 GHz IEEE 802.15.4 / ZigBee-ready RF Transceiver, http://inst.eecs.berkeley.edu/~cs150/Documents/CC2420.pdf

\bibitem{francesco} M.~D.~Francesco, G.~Anastasi, M.~Conti, S.~K.~Das, and V.~Neri, ``Reliability and Energy-Efficiency in IEEE 802.15.4/ZigBee Sensor Networks: An Adaptive and Cross-Layer Approach,''  {\em IEEE Journal on Selected Areas in Communications}, 29(8), pp. 1508--1524, September 2011.

\bibitem{Gilani} M.~H.~S.~Gilani, I.~Sarrafi, M.~Abbaspour, ``An Adaptive CSMA/TDMA Hybrid MAC for Energy and Throughput Improvement of Wireless Sensor Networks,'' {\em Ad Hoc Networks (2011)}, pp. 1--8, 2011.

\bibitem{Zhuo} S.~Zhuo, Y.~Song, Z.~Wang, and Z.~Wang, ``Queue-MAC: A Queue-length Aware Hybrid CSMA/TDMA MAC Protocol for Providing Dynamic Adaptation to Traffic and Duty-cycle Variation in Wireless Sensor Networks,''  in {\em Proc. of 9th IEEE International Workshop on Factory Communication Systems (WFCS),} pp.105--114, 21-24 May 2012.

\bibitem{Ephrem} A.~Ephremides, O.~A.~Mowafi, ``Analysis of a Hybrid Access Scheme for Buffered Users-Probabilistic Time Division,'' {\em IEEE Transactions on Software Engineering,} vol.~SE-8, no.1, pp. 52--61, January 1982.

\bibitem{cheng} S.~T.~Cheng and M.~Wu, ``Contention-Polling Duality Coordination Function for IEEE 802.11 WLAN Family,'' {\em IEEE Transactions on  Communications}, 57(3), pp.~779--788, March 2009.

\bibitem{bin}
B.~Liu, Z.~Yan, and C.~W.~Chen, ``CA-MAC: A Hybrid Context-Aware MAC Protocol for Wireless Body Area Networks,'' in {\em Proc. of 13th IEEE International Conference on e-Health Networking Applications and Services (Healthcom)}, pp.~213--216, 13-15 June 2011.

\bibitem{zmac} I.~Rhee, A.~Warrier, M.~Aia, J.~Min, and M.~L,~Sichitiu, ``Z-MAC: A Hybrid MAC for Wireless Sensor Networks,'' {\em IEEE/ACM Transactions on Networking,}  vol.~16, no.~3, pp.~511--524, June 2008.

\bibitem{Wei} W.~Wang, H.~Wang, D.~Peng, and H.~Sharif, ``An Energy Efficient Pre-Schedule Scheme for Hybrid CSMA/TDMA MAC in Wireless Sensor Networks,'' in {\em Proc. of 10th IEEE Singapore International Conference on Communication Systems}, pp.~1--5, October 2006.

\bibitem{Rana} Y.~K.~Rana, B. H. Liu, A.~Nyandoro, and S.~Jha, ``Bandwidth Aware Slot Allocation in Hybrid MAC,''  in {\em  Proc. of 31st IEEE Conference on Local Computer Networks,} pp.~89--96, 14-16 November 2006.

\bibitem{sheu} S.~T.~Sheu, Y.~Y.~Shih, and W.~T.~Lee, ``CSMA/CF Protocol for IEEE 802.15.4 WPANs,'' {\em IEEE Transactions on Vehicular Technology}, 58(3), pp.~1501--1516, March 2009.

\bibitem{Zhang2} R.~Zhang, L.~Cai, J.~Pan, ``Performance Study of Hybrid MAC Using Soft Reservation for Wireless Networks,'' in {\em Proc. of 2011 IEEE International Conference on Communications (ICC)}, pp.~1--5, 5-9 June 2011.

\bibitem{jurdak} R.~Jurdak, A.~G.~Ruzzelli, and G.~M.~P.~O'Hare, ``Radio Sleep Mode Optimization in Wireless Sensor Networks,'' {\em IEEE Transactions on Mobile Computing}, 9(7), pp.~955--968, July 2010.

\bibitem{khanafer} M.~Khanafer, M.~Guennoun, and H.~T.~Mouftah, ``Adaptive Sleeping Periods in IEEE 802.15.4 for Efficient Energy Savings: Markov-Based Theoretical Analysis,'' in {\em Proc. of IEEE ICC'11}, pp.~1--6, Japan, June 2011.

\bibitem{xiao} Z.~Xiao, C.~He, and L.~Jiang, ``Slot-based Model for IEEE 802.15.4 MAC with Sleep Mechanism,'' {\em IEEE Communications Letters}, 14(2), pp.~154--156, February 2010.

\bibitem{gts} B.~Shrestha, E.~Hossain, S.~Camorlinga, R.~Krishnamoorthy, and D.~Niyato, ``An Optimization-based GTS Allocation Scheme for IEEE 802.15.4 MAC with Application to Wireless Body Area Sensor Networks," in {\em Proc. of IEEE ICC'10}, pp.~1--6, Cape Town, May 2010.

\bibitem{Sikdar} A.~Seyedi and B.~Sikdar, ``Energy Efficient Transmission Strategies for Body Sensor Networks with Energy Harvesting,''  { \em IEEE Transactions on Communications}, 58(7), pp.~2116--2126, July 2010.

\bibitem{Liu}
 Z.~Liu and I.~Elhanany, ``RL-MAC: A Reinforcement Learning Based MAC Protocol for Wireless Sensor Networks,'' {\em International Journal of Sensor Networks (IJSNET)}, 1(3/4), pp.~117--124, April 2006.

\bibitem{Angel}
G.~d.~Angel and T.~L.~Fine, ``Optimal Power and Retransmission Control Policies for Random Access Systems,'' {\em IEEE/ACM Transactions on Networking}, 12(6), pp.~1156--1166, December 2004.

\bibitem{Phan}
C.~V.~Phan, Y.~Park, H.~H.~Choi, J.~Cho, and J.~G.~Kim, ``An Energy-Efficient Transmission Strategy for Wireless Sensor Networks,'' {\em IEEE Transactions on Consumer Electronics}, 56(2), pp.~597--605, May 2010.

\bibitem{van}
 N.~Mastronarde and M.~van der Schaar, ``Fast Reinforcement Learning for Energy-Efficient Wireless Communications,'' {\em IEEE Transactions on Signal Processing}, 59(12), pp.~6262--6266, December 2011.


\end{thebibliography}
\end{document}